\documentclass[11pt]{article}
\usepackage[british]{babel}
\usepackage{float}
\usepackage{pifont}
\usepackage{rotating}
\usepackage[normalem]{ulem}
\usepackage[table]{xcolor}
\usepackage{tikz}
\usepackage{amsmath,amssymb,amsmath,amsthm,epsfig,graphicx}
\usepackage{grffile}
\usepackage[round]{natbib}
\usepackage{color}
\usepackage{appendix}
\usepackage{enumitem}
\usepackage{authblk}
\usepackage{lscape}
\usepackage{subcaption}
\usepackage{bm}
\usepackage{lineno}
\usepackage{algorithm}
\usepackage{algorithmicx}
\usepackage{algpseudocode}
\usepackage{mwe}
\usepackage{colortbl}
\usepackage{geometry}
\usepackage{mathrsfs}
\usepackage[short,c2,nocomma]{optidef}

\usetikzlibrary{arrows,%
                shapes,positioning,calc, patterns}
\RequirePackage{ifpdf}
\ifpdf
	\usepackage{graphicx}	
	\DeclareGraphicsExtensions{.jpg, -eps-converted-to.pdf}
\else
	\usepackage{graphicx}		
	\DeclareGraphicsExtensions{-eps-converted-to.pdf}
\fi

\usepackage{hyperref}
\hypersetup{%
  colorlinks=true,
  linkcolor=blue,
  citecolor=blue,
}

\newcommand{\Keywords}[1]{\par\noindent
{\small{\bf Keywords\/}: #1}}

\begin{document}

\title
{From snapshots to manifolds - A tale of shear flows}

\author[1]{E. Farzamnik \thanks{ehsan.farzamnik@gmail.com}}
\author[2]{A. Ianiro \thanks{aianiro@ing.uc3m.es}}
\author[2]{S. Discetti}
\author[3,4]{N. Deng}
\author[5]{K. Oberleithner}
\author[3,6]{B.~R.~Noack}
\author[1]{V. Guerrero}
\affil[1]{\small Department of Statistics, Universidad Carlos III de Madrid,  Spain \vspace*{0.25cm}}
\affil[2]{\small Aerospace Engineering Research Group, Universidad Carlos III de Madrid,  Spain \vspace*{0.25cm}}
\affil[3]{\small School of Mechanical Engineering and Automation
Harbin Institute of Technology, \newline Shenzhen, P.~R.~China \vspace*{0.25cm}}
\affil[4]{\small Institute of Mechanical Sciences and Industrial Applications, ENSTA-Paris,  France \vspace*{0.25cm}}
\affil[5]{\small Laboratory for Flow Instability and Dynamics, Technische Universität Berlin,  Germany \vspace*{0.25cm}}
\affil[6]{\small Hermann-Föttinger-Institut, Technische Universität Berlin,  Germany \vspace*{0.25cm}}

\date{}

\maketitle

\begin{abstract}
We propose a novel non-linear manifold learning from snapshot data
and demonstra\-te its superiority over Proper Orthogonal Decomposition (POD) for shedding-dominated shear flows.
Key enablers are isometric feature mapping, Isomap \citep{Tenenbaum2319},
as encoder and  $K-$nearest neighbours ($K$NN) algorithm as decoder.

The proposed technique is applied to numerical and experimental datasets including the fluidic pinball, a swirling jet, and the wake behind a couple of tandem cylinders. 
Analyzing the fluidic pinball, the manifold is able to describe the pitchfork bifurcation and the chaotic regime with only three feature coordinates.  
These coordinates  are linked to  vortex-shedding phases and the force coefficients. 
The manifold coordinates of the swirling jet are comparable to the POD mode amplitudes,
yet allow for a more distinct manifold identification 
which is less sensitive to measurement noise.
As similar observation is made for the wake of two tandem cylinders \citep{RAIOLA2016354}. 
The tandem cylinders are aligned in streamwise distance 
which corresponds to the transition between the single bluff body 
and the reattachment regimes of vortex shedding. 
Isomap unveils these two shedding regimes while 
the Lissajous plots of first two POD mode amplitudes feature a single circle.

The reconstruction error of the manifold model 
is small compared to the fluctuation level,
indicating that the low embedding dimensions contains the coherent structure dynamics. 
The proposed Isomap-$K$NN manifold learner is expected to be of large importance
in estimation, dynamic modeling and control for  large range of configurations with dominant coherent structures.

\vspace*{0.3cm}
\end{abstract}

\section{Introduction}
\label{sec:intro}
The chaotic nature of turbulent flows and their importance in physical and engineering systems has motivated countless studies attempting to obtain simplified models for control purposes in engineering applications. In particular, unbounded shear flows such as jets and wakes have received unabated interest due to their importance for drag reduction, control of unsteady loads, mixing enhancement etc. \par

Despite their chaotic nature, such flows are characterized by recurrent flow patterns that are typically referred to as coherent structures. The beauty of the coherent structures has fascinated scientists since Leonardo Da Vinci \citep{leonardo2021} and has hinted at the possibility that the flow dynamics can be represented as a system evolving on a low-dimensional attractor.
Proper Orthogonal Decomposition  \citep[POD,][]{BerkoozPOD1993}, also called Principal Component Analysis in statistics, has received significant attention since it allows to decompose a flow field into orthogonal modes sorted according to their contribution to the variance of the quantity to be analyzed. One of the most exploited advantages of POD in fluid mechanics is their capability to simplify the Navier-Stokes equations into a system of linear differential equations employing Galerkin projections \citep{noack2003hierarchy}.
Low-order models obtained from POD open a door to a vast space of applications such as flow control \citep{brunton2015closed} and also crisp least-order models for bifurcations and interactions of coherent structures \citep{Deng_2019}.

A myriad of POD studies hint at low-dimensional manifolds 
describing turbulent shear flows.
In the case of oscillatory flows, 
two-dimensional manifolds have been 
identified from laminar two-dimensional cylinder flows \citep{noack2003hierarchy} 
to experimental turbulent wakes behind finite cylinders at high Reynolds numbers \citep{Bourgeois2013jfm}.
These manifolds are the cornerstone of mean-field Galerkin models.
Even flows with several frequencies 
may live on a mean-field manifold  \citep{Luchtenburg2009jfm}.
The pioneering POD model of \cite{Aubry1988jfm} derives such a manifold for the turbulent boundary layer from the Reynolds equations.
For more complex flows, the energy spectrum of POD
typically reveals $O(10)$ distinct most energetic eigenvalues
associated with physically interpretable modes.
This distinct spectrum is followed by a continuous eigenvalue distribution
with less interpretable and increasingly fine-scaled modes.
The distinct POD modes can be hypothesized as the conductors
of a large turbulence orchestra `slaved' 
their conductor POD mode amplitudes \citep{Callaham2022arXiv}.

While the focus of POD is on obtaining reduced-order models optimal in terms of energy, the field of statistical learning provides a vast amount of tools for dimensionality reduction \citep{franklin2005elements}. Multi-dimensional scaling (MDS), for instance, is based on the singular value decomposition of the data distance matrix and allows representing a dataset in a low-dimensional space preserving the distance between the snapshots in the high-dimensional space \citep{Torgerson1952, kruskal1964nonmetric}. MDS has been used in fluid mechanics mainly for visualization purposes and effectively captures some hidden features of the flows \citep{kaiser2014,Foroozan2021}.

The capability of obtaining low-dimensional representations is tempting to identify embedded manifolds of the flows under study. 
Manifold learning attempts to identify a topologically closed surface, the manifold, over which the dataset actually resides. In statistical words, the dataset can be described to lie on or near a manifold in a low-dimensional space in which the manifold expresses some basic features of it. In this sense the manifold is in fact the set of relations that connect snapshots to each other. 
Interestingly, often high-dimensional systems appear to evolve on low-dimensional manifolds, thus simpli\-fying their modelling if the manifold can be identified. These aspects pushed the development of the manifold learning techniques in the last decades. 
Remarkable examples are locally linear embedding \citep{roweis2000nonlinear}, isometric mapping \citep{Tenenbaum2319}, and diffusion map \citep{coifman2006diffusion}.

The traditional dimensionality-reduction methods are structured on linear models and thus fail in capturing the manifolds when a non-linear structure is present in the data. 
Turbulent flows exhibit a non-linear behaviour which motivates the investigation of non-linear models for manifold learning. \cite{Tenenbaum2319} have shown that the dimen\-sionality reduction based on geodetic distances can be a powerful tool in preserving the actual behaviour of non-linear datasets. This technique is referred to as isometric feature mapping, or Isomap. Surprisingly, the application of Isomap in fluid mechanics is minimal. \cite{Tauro2014} successfully employed Isomap to identify manifolds from flow-visualization data while others used it for combustion \citep{bansal2011identification}, and design optimization \citep{Gortz2014}. Recently \cite{otto2021inadequacy} discussed the limitation of the linear methods in the case of selection and placement of sensors in a flow field. \par

This manuscript introduces a framework of manifold learning as an encoder for unbounded shear flows with a $K-$Nearest Neighbors ($K$NN) decoder. The input snapshots, which can be obtained either from a simulation or an experiment, are encoded using Isomap as the primary tool. The high-dimensional space transforms to a low-dimensional space to identify the hidden embedding manifold of the dataset. In this new space, the manifold is interpreted to unravel the relationship between the manifold low-dimensional characteristics and the main features of the flow dynamics. We can reproduce the snapshots in the high-dimensional space with a $K$NN decoder using this new, easy-to-understand space and fast computing. This whole encoder-decoder model provides a robust framework to analyze shear flows and then implement applications (such as designing flow control systems) based on it. \par

Four datasets with different features have been used to investigate the framework's performance. The selected datasets vary from DNS simulations of wake flows to fully-turbulent experimental datasets with measurement noise. The simulation datasets are based on the wake of the fluidic pinball which in recent years has been shown to be a suitable test-bed configuration to study general flow phenomena like bifurcations and flow control \citep{Deng_2019}. To study different flow regimes, the results from the simulations at $Re = 80$ and $Re = 130$ are reported and discussed, allowing to identify the manifold learning capabilities both in a simpler bifurcation and in a more complex chaotic environment. 
The first experimental dataset consists of PIV measurements in a highly functional swirling jet configuration. This configuration has a wide application in modern gas turbine combustors and aerodynamically stabilizes lean premixed flames \citep{swirling_jet,luckoff2021mean}. Both the turbulent regime and the measurement noise challenge the encoder-decoder framework. The last tested dataset relates to the flow in the wake of two tandem cylinders. Tandem cylinders are characterized by several working regimes depending on the streamwise cylinder distance. The proposed dataset is at the intersection of two regimes, however by using POD in a previous work, \cite{RAIOLA2016354} could not unveil a regime switch. \par

The paper is organized as follows: in \S \ref{sec:dim_red_methods} after the introduction, a detailed description of the developed framework is provided; the datasets and flow configurations employed are described in \S \ref{sec:data}; the most important outcomes from the analysis using the encoder-decoder framework are presented in \S \ref{sec:results}, and finally, the conclusion and the possible future steps have been put in \S \ref{sec:conclusion}. Two appendices describe a criterion for the choice of Isomap parameters and discuss possible criteria for the definition of the manifold residual variance.
\section{Isomap - $K$NN manifold learner}
\label{sec:dim_red_methods}

 \begin{figure}
  \centerline{\includegraphics[width=1\textwidth]{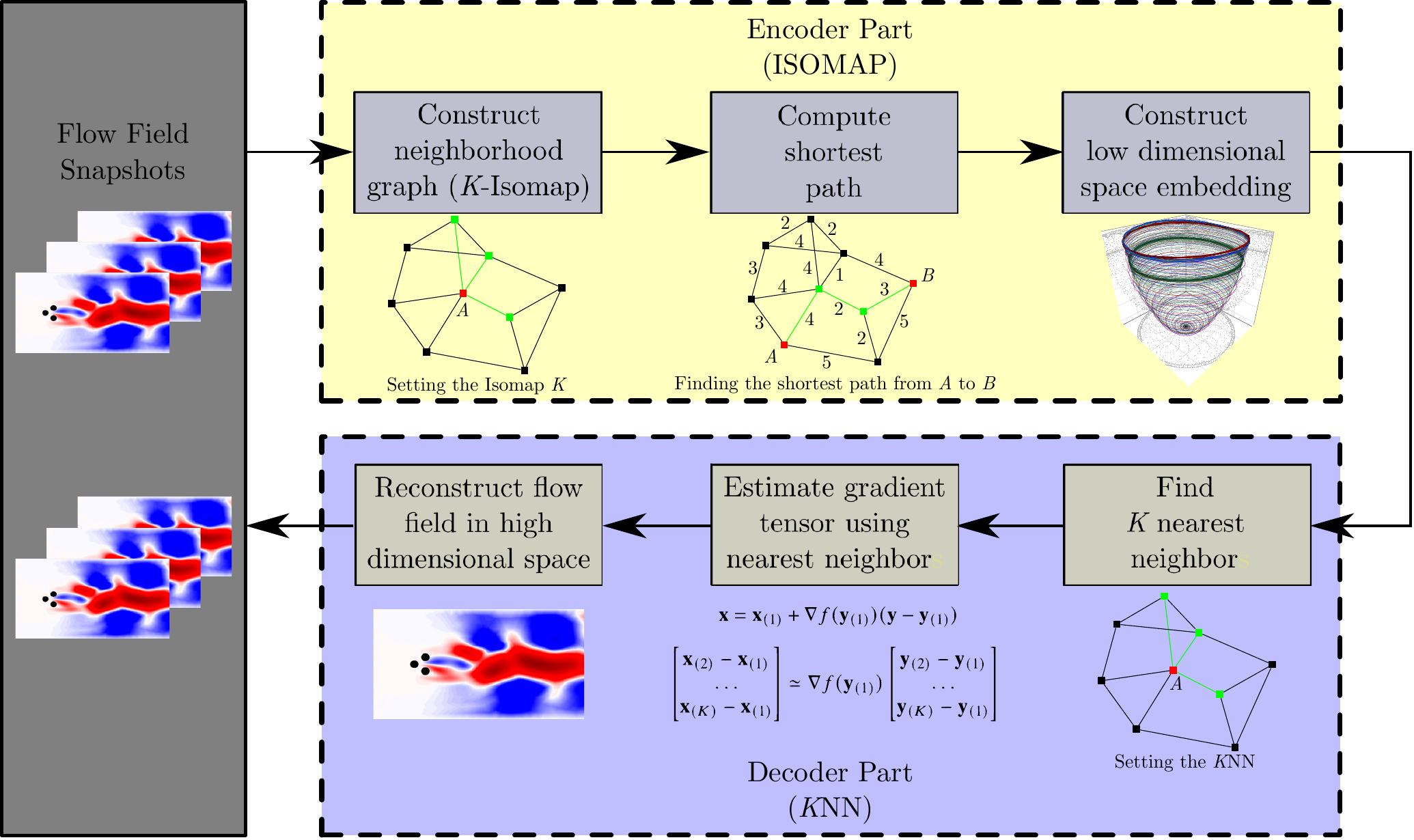}}
  \caption{Encoder-Decoder procedure. Left block: obtaining flow field snapshots from DNS or experiment- Top block: encoder part, representing the Isomap method application on input snapshots to identify the embedding manifold in low-dimensional space- Bottom block: decoder part which reconstruct the flow field snapshots from low-dimensional space coordinates.}
\label{fig:flowchart}
\end{figure}
 
In this work, a manifold learner methodology for fluid data is developed. The proposed approach consists of three steps. First, data is gathered either from simulations or experiments.
Second, the so-obtained data is embedded into a low-dimensional space using Isometric feature Mapping \citep[Isomap,][]{Tenenbaum2319}. This encoding part, which is fully data-driven, is carried out with the aim of revealing a hidden manifold that allows us to relate the new coordinates to physical features of the flow such as, for instance, force coefficients.
Finally, a decoding part that enables return to the high-dimensional space and reconstructs the original flow field is developed. The proposed decoder is based on  $K-$nearest neighbours ($K$NN) and linear interpolation. Figure \ref{fig:flowchart} shows the three stages of our procedure, which are described in detail in what follows.

Let us consider that $N$ flow field snapshots have been observed, either from an experimental setting or a simulation. Then, each snapshot is an observation (point) in the high-dimensional space $\mathbb{R}^P$, where each dimension (feature) contains information about a point of the field. Let $\mathbf{X}\in{\mathbb R}^{N\times P}$ be the data matrix containing the stated information and $\mathbf{x}_i\in \mathbb{R}^P$ be each of its rows, i.e.\ the flow fields for $i=1,\ldots,N.$
The dataset in $\mathbf{X}$  is complex by nature and being able to extract a meaningful small number of coordinates that capture the main characteristics of the flow is challenging.

Isomap is a nonlinear dimensionality reduction technique that finds a low-dimensional embedding of the data points that best preserve the geodesic distances measured in the high-dimensional input space. In order to estimate these geodesic distances, the shortest paths in a graph connecting neighbouring points are employed. These distances are then used as an input in classical MDS \citep{Torgerson1952} to construct the low-dimensional embedding so that the Euclidean pairwise distances resemble those in the neighbouring graph. Therefore, the  Isomap algorithm runs as follows. First,  the Euclidean distances  $d_{\mathbf{X}}(i,j)$ between flow fields $\mathbf{x}_i$  and $\mathbf{x}_j$ are computed for all $i,j=1,\ldots,N$. Second, for
$i=1,\ldots,N$ , $\mathcal{N}_{\mathbf{X}}^k(i),$ is defined as the set of the $k$ closest observations to $\mathbf{x}_i.$ Based on these neighborhoods, the neighboring graph $G$ is defined over these data points such that two nodes (flow fields) $i$ and $j$ are connected by an edge of weight $d_{\mathbf{X}}(i,j)$ if they are neighbors, i.e.\ there is an edge between $i$ and $j$ if $\mathbf{x}_j\in \mathcal{N}_{\mathbf{X}}(i)$. Observe that $G$ approximates the high-dimensional manifold containing the observed data. Third, the shortest paths between all pair of vertices in $G$ are computed, yielding $d_{\mathbf{G}}(i,j)$ for all $i,j=1,\ldots,N$, using Floyd's algorithm \citep{Floyd1962}. Let $\mathbf{D}_G$ be the matrix containing these shortest path distances. Finally, obtain the low-dimensional embedding $\mathbf{\Gamma}\in{\mathbb R}^{N\times p},$  $p<<P$, using MDS. The new coordinates for the $N$ samples are then found so that their pairwise Euclidean distance resembles $d_{\mathbf{G}}(i,j)$. This is equivalent to finding $\mathbf{\Gamma}$ which minimizes the cost function

\begin{equation}\label{eq:classicalMDS}
\left\Vert \mathbf{\Gamma\Gamma}^\top - \mathbf{B}\right\Vert_F^2,
\end{equation}
where $\mathbf{B} = -\displaystyle\frac{1}{2}\mathbf{H}^\top \left(\mathbf{D}_G \odot \mathbf{D}_G\right)\mathbf{H}$ is the Gram matrix in the input space, being  $\mathbf{H}= \mathbf{I}_N - \displaystyle\frac{1}{N}\mathbf{I}_N$ the centering matrix, $\mathbf{I}_N$ the identity matrix of dimension $N$, $\odot$ the Hadamard (element-wise) product and $\|\cdot\|_F$ the Frobenius norm.  

Fixing a dimension $p$ for the low-dimensional embedding, the value of $\mathbf{\Gamma}$ minimizing the quantity in \eqref{eq:classicalMDS} is the matrix made up of the $p$ eigenvectors $\boldsymbol{\gamma}_1,\ldots,\boldsymbol{\gamma}_p$ corresponding to the $p$ largest (positive) eigenvalues of the matrix $\mathbf{\Lambda}$ arising from the eigendecomposition of $\mathbf{B}$, namely $\mathbf{B} = \mathbf{V\Lambda V}^\top$ and $\mathbf{\Gamma} = \mathbf{V}_p.$

The aforementioned Isomap algorithm  admits the choice of other norms different from the Euclidean, different ways of identifying the neighbors to construct $G$, other shortest path algorithms or a non-classical approach to MDS. However, the choices made in our methodology are motivated by the implemented version of Isomap in the {\tt RDRToolbox} in the R software \citep{Rsoftware,BartenhagenR}, which has been used to carry out our analyses.

In order to assess the performance of Isomap, \cite{Tenenbaum2319} propose using the  definition of {\it residual variance} as in \eqref{eq:residualvarianceT}. Let $\mathbf{D}_{\mathbf{\Gamma}}$ be the matrix of Euclidean distances between each pair of points in the low-dimensional embedding. Then, the residual variance is defined as one minus the squared correlation coefficient between the vectorization of the distance matrices $ \mathbf{D}_G$ and $\mathbf{D}_{\mathbf{\Gamma}}$, yielding

\begin{equation}\label{eq:residualvarianceT}
1- R^2(\text{vec}\left(\mathbf{D}_G\right), \text{vec}\left( \mathbf{D}_{\mathbf{\Gamma}}\right)),
\end{equation}
where $R^2$ refers to the squared correlation coefficient and vec is the vectorization operator. Observe that the result in \eqref{eq:residualvarianceT} is a number between $0$ and $1$ which  accounts for the amount of information that remains unexplained by the low-dimensional embedding of the original data. Therefore, the lower the value in \eqref{eq:residualvarianceT} the better. For a discussion about the definition of residual variance to assess the performance of Isomap against other dimensionality reduction methods such as POD we refer the reader to Appendix \ref{app:residualvariance}.

In order to provide a decoder to create a correspondence between Isomap coordinates $\boldsymbol{\gamma}_1,\ldots,\boldsymbol{\gamma}_p$ and the ones in the high-dimensional space, namely $\mathbb{R}^P,$  
we employ a purely data-driven approach.
Any flow field $\mathbf{x}_i\in\mathbb{R}^P$ has its low-dimensional counterpart $\mathbf{y}_i~\in~\mathbb{R}^p,\, i=1,\ldots,N.$ Then, let $f:\mathbb{R}^p\longrightarrow\mathbb{R}^P$ be the unknown mapping which transforms the flow fields in the low-dimensional space onto the high-dimensional ones. To reconstruct the flow field for any $\mathbf{y}\in\mathbb{R}^p$ we assume that
its  $K-$Nearest Neighbors  $\mathbf{y}_{(1)},\ldots, \mathbf{y}_{(K)}$  and their high-dimensional counterparts, namely $\mathbf{x}_{(1)},\ldots, \mathbf{x}_{(K)},$ are identified. Therefore, the reconstruction (or decoding) of $\mathbf{y},$ denoted as $\mathbf{x},$ can be obtained as a first order Taylor expansion starting from the nearest neighbor to be mapped back to the original space, i.e.\ $\mathbf{x}_{(1)},$ as

\begin{equation}
    \mathbf{x} = \mathbf{x}_{(1)} + \nabla f(\mathbf{y}_{(1)}) (\mathbf{y} - \mathbf{y}_{(1)}), 
\end{equation}

\noindent where the gradient tensor in $\mathbf{y}_{(1)}$, namely $\nabla f(\mathbf{y}_{(1)})=\left(\frac{\partial f}{\partial \boldsymbol{\gamma_1}}(\mathbf{y}_{(1)}),\ldots, \frac{\partial f}{\partial \boldsymbol{\gamma_p}}(\mathbf{y}_{(1)})\right)$ is estimated assuming an orthogonal projection of the $K-1$ directions provided by the $K-$Nearest Neighbors in $\mathbb{R}^P$ to those in $\mathbb{R}^p.$ This is:

\begin{equation}\label{eq:knnproj}
\begin{bmatrix}
\mathbf{x}_{(2)} -  \mathbf{x}_{(1)}\\
\ldots \\
\mathbf{x}_{(K)} -  \mathbf{x}_{(1)}
\end{bmatrix}
\simeq
\nabla f(\mathbf{y}_{(1)})
\begin{bmatrix}
\mathbf{y}_{(2)} -  \mathbf{y}_{(1)}\\
\ldots \\
\mathbf{y}_{(K)} -  \mathbf{y}_{(1)}
\end{bmatrix},
\end{equation}
which yields $\nabla f(\mathbf{y}_{(1)}) = (\Delta \mathbf{Y}^\top \Delta\mathbf{Y})^{-1}\Delta\mathbf Y^\top \Delta \mathbf{X}$ if least squares minimization is used to approximate it, and  $\Delta \mathbf{X}$ and $\Delta \mathbf{Y}$ are the left-hand side and the second term in the right-hand side of \eqref{eq:knnproj}, respectively.

\section{Datasets \label{sec:data}}

In this section we describe the datasets that have been used to test our methodology. Three configurations are considered, which yield different flow fields and regimes under both experimental and simulation setups.

\begin{figure}
     \centering
          \begin{subfigure}[b]{0.49\textwidth}
         \includegraphics[width=\textwidth]{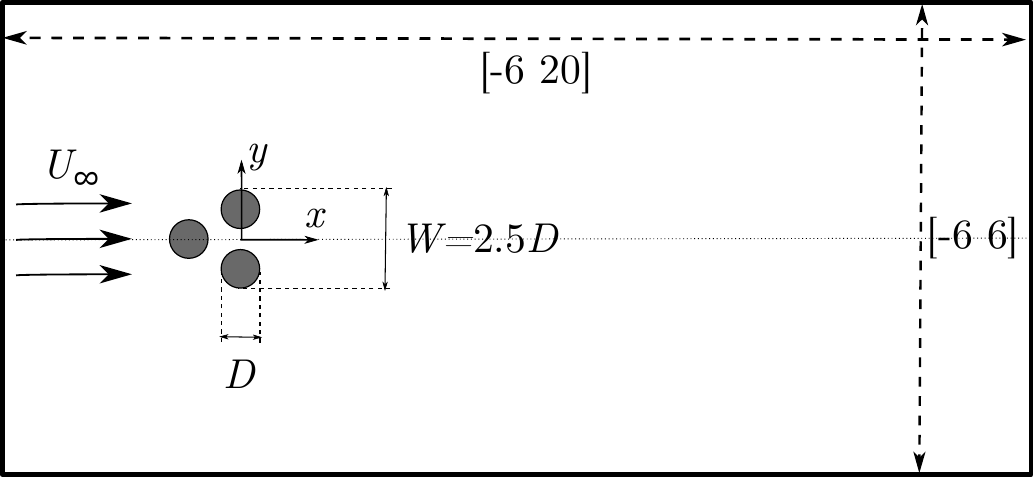}
         \caption{}
         \label{fig:pinball_conf}
     \end{subfigure}
     \hfill
     \begin{subfigure}[b]{0.49\textwidth}
         \includegraphics[width=\textwidth]{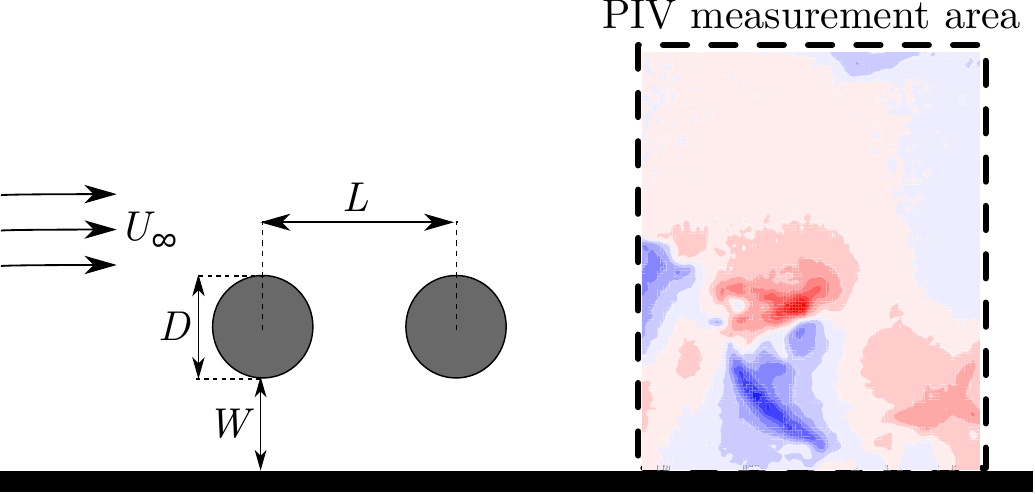}
         \caption{}
         \label{fig:tandem_conf_contours}
     \end{subfigure}
          \hfill
     \begin{subfigure}[b]{0.49\textwidth}
     \includegraphics[width=\textwidth]{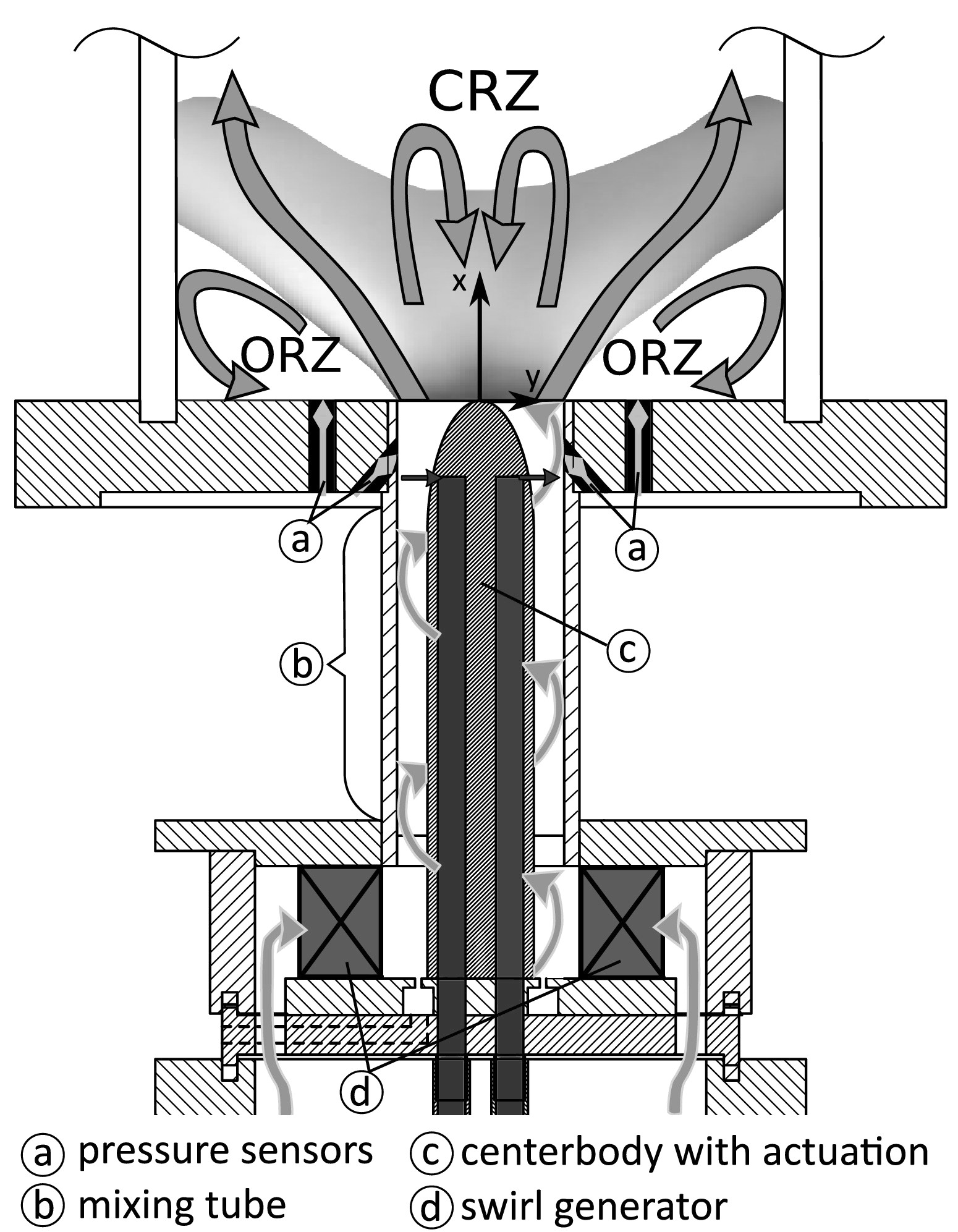}
     \caption{}
     \label{fig:swirlingjet_conf_contours}
     \end{subfigure}
        \caption{Dataset Configurations. a) Fluidic Pinball- b) Tandem Cylinders- c) Swirling Jet (Reprinted from \cite{luckoff2021mean} with permission from Elsevier).}
        \label{fig:configurations}
\end{figure}

\subsection{Fluidic pinball dataset}
The fluidic pinball is a flow configuration consisting of three rotatable cylinders of equal diameter $D$ whose axes are located in the vertices of an equilateral triangle, as sketched in figure \ref{fig:pinball_conf}.
The triangle set of three cylinders has a center-to-center side length $3D/2$ and is immersed in a viscous incompressible flow with a uniform upstream velocity $U_\infty$.
The Reynolds number for this setup is defined as $Re = U_\infty D/\nu$, where $\nu$ is the kinematic viscosity of the fluid.
The wake flow undergoes a set of interesting transitions at different values of the Reynolds number, thus allowing to explore reduced-order modeling and flow control strategies in a wide range of scenarios.
In the recently published literature, the fluidic pinball has been used as a benchmark configuration for testing the mean-field modeling \citep{Deng_2019}, cluster-based network modeling \citep{deng_2022}, and machine learning control \citep{cornejo2021jfm, doi:10.1063/1.5127202}.

\begin{figure}
     \centering
     \begin{subfigure}[b]{0.49\textwidth}
         \includegraphics[width=\textwidth]{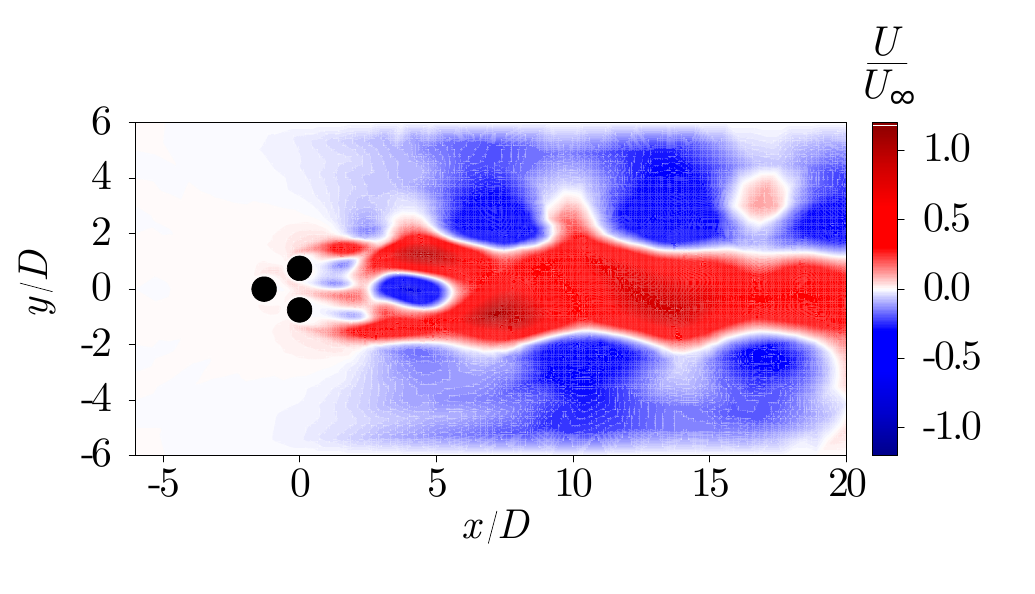}
         \caption{}
         \label{fig:Re80_u_flow_field}
     \end{subfigure}
     \hfill
     \begin{subfigure}[b]{0.49\textwidth}
         \includegraphics[width=\textwidth]{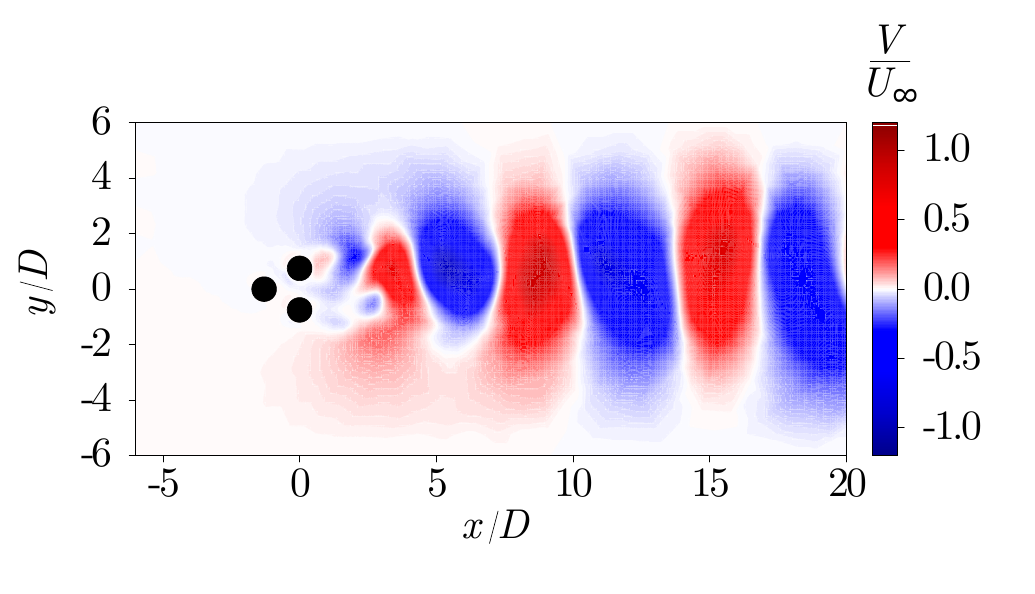}
         \caption{}
         \label{fig:Re80_v_flow_field}
     \end{subfigure}
          \hfill
     \begin{subfigure}[b]{1\textwidth}
         \includegraphics[width=\textwidth]{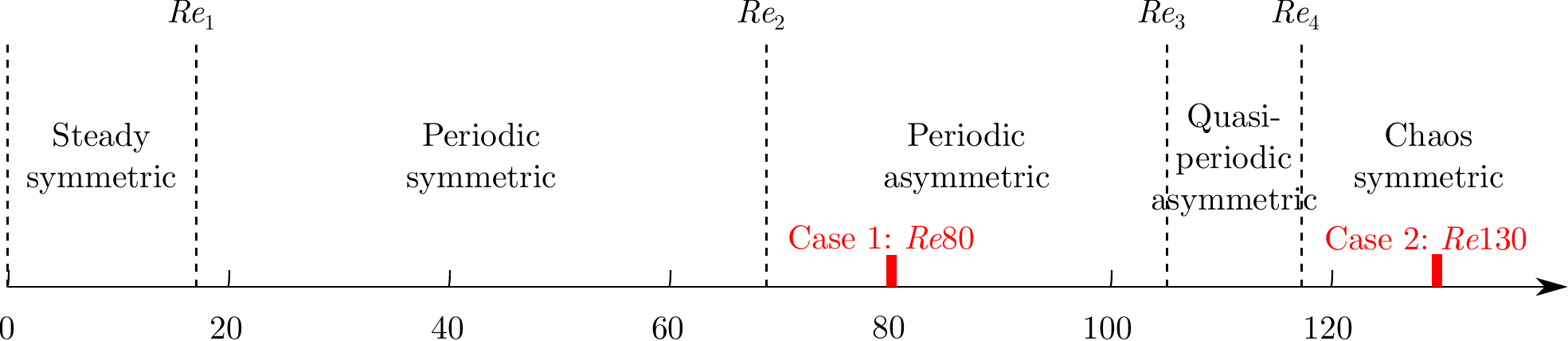}
         \caption{}
         \label{fig:pinball_regimes}
     \end{subfigure}
        \caption{An example snapshot of the fluidic pinball after subtracting the steady solution: 
        a) the contour of the streamwise fluctuating velocity component, and 
        b) the contour of the crosswise fluctuating velocity component, normalized with the upstream velocity $U_\infty$.
        c) Transitions of different flow regimes with varying Reynolds numbers.}
        \label{fig:pinball_conf_contours}
\end{figure}

The numerical results of the fluidic pinball are investigated by employing a software developed to study multiple-input multiple-output flow control by \citet{pinball_simulator}.
Direct numerical simulation of the incompressible Navier-Stokes equations is used to compute the two-dimensional viscous wake behind the pinball configuration.
As shown in figure \ref{fig:pinball_conf_contours}(a), the computational domain $[-6D,20D] \times [-6D,6D]$, excluding the interior of the cylinders, is described by a Cartesian coordinate system whose origin is located in the middle of the top and bottom cylinders.
As the rotation of the cylinders is not considered in this study, a no-slip condition on the cylinders and the far-field velocity $U_\infty$ are considered as the boundary conditions.
The unsteady Navier-Stokes solver is based on fully implicit time integration using an iterative Newton-Raphson approach and finite-element method (FEM) discretization on an irregular grid structure with $4225$ triangles and $8633$ vertices \citep{Deng_2019}.
The steady solution, used as the initial condition at each corresponding Reynolds number, is calculated by the solver for the steady Navier-Stokes equations  in the same way.

With increasing Reynolds number, the flow experiences a transition from a laminar flow to periodic vortex shedding and finally to chaos \citep{Deng_2019}.
Five different regimes have been identified, as summarized in figure \ref{fig:pinball_regimes}.
The transition from a steady symmetric flow to a periodic symmetric vortex shedding occurs at $Re_1 \approx 18$ \citep[following a Hopf bifurcation,][]{andronov1971theory, Strogatz1994Hoptbifurcation}.
The symmetry of the vortex shedding vanishes at $Re_2 \approx 68$ \citep[pitchfork bifurcation,][]{Strogatz1994Hoptbifurcation} thus entering a periodic asymmetric regime.
A secondary frequency appears with a higher Reynolds number, and the flow experiences another transition to a quasi-periodic asymmetric regime at $Re_3 \approx 104$ \citep[Neimark–Säcker bifurcation,][]{kuznetsov2008neimark}. 
Finally, at $Re \approx 115$, the flow enters into a chaotic symmetric regime. 

In this study, we focused on two different flow states at the selected $Re$, representative of the two most complex flow states identified by \cite{deng_2022}.
At $Re=80$ for the periodic asymmetric regime, there exist a total of six invariant sets in the system state space: three unstable stable solutions, one unstable limit cycle, and two stable limit cycles, resulting from the primary Hopf bifurcation and the secondary pitchfork bifurcation. 
The dataset at $Re=80$ includes the snapshots from the simulations starting from the symmetric steady solution and its mirror-conjugated snapshots, as well as the snapshots from the simulations starting from the two mirror-conjugated asymmetric steady solutions.
In this case, it is able to ensure that the dataset contains the complete manifold with six invariant sets.
At $Re=130$ for the chaotic symmetric regime, three unstable stable solutions and one chaotic attracting set can be found.
The dataset considered at $Re=130$ only contains the snapshots from the simulation starting from the symmetric steady solution, since we are interested in the transition from the unstable invariant set to the chaotic attracting set in this case.
Therefore, simulations are performed at $Re$ equal to $80$ and $130$, thus covering the typical complex wake dynamics regime with multiple invariant sets and chaotic attracting set.

\subsection{Swirling jet dataset}

Swirling jets have a wide variety of applications in modern gas turbine combustors and aerodynamically stabilize lean premixed flames. In present work we analyze an experimental dataset obtained with stereoscopic particle image velocimetry by \cite{luckoff2021mean}.
Figure \ref{fig:swirlingjet_conf_contours} reports a schematic of the swirling nozzle and of the measurement domain employed in the work by \cite{luckoff2021mean}. This configuration consists of a feeding line which provides the mass flow rate to a jet issued from a swirling nozzle. 
The swirling flow is produced using a radial swirl generator and the swirl number $Sw$, defined as the ratio of the axial flux of tangential momentum to the axial flux of axial momentum, can range between 0 and 1.5. The nozzle exit has a diameter of $D = 55mm$ and has a centerbody at the center with diameter $D_{CB} = 35mm$ thus the hydraulic diameter of the mixing tube has a diameter of $D_h = 20mm$. The Reynolds number is defined using the hydraulic diameter and the experimental facility can provide jets with a Reynolds number in the range $[13000, 32000]$. Present dataset has been generated for $Re = 20000$ and $Sw = 0.7$. 

The measurement domain is located at the nozzle exit to analyze the flow in the combustion chamber. 
The combustion chamber is a cylinder with a inner diameter of $D_{CCh} = 200mm$ and length of $L_{CCh} = 300mm$ and is made of quartz glass to enable flow measurement using time-resolved particle image velocimetry (PIV). For the present dataset, 2183 snapshots were captured and evaluated using a commercial PIV software. The raw data were filtered for removing the outliers. The velocity at the actuators outlet has been measured using hot-wire sensors. The reader can refer to the work by \cite{luckoff2021mean} for further details on the measurement setup and PIV image processing. \par

The main coherent structure in this kind of flow is a helical structure known as the precessing vortex core \citep[PVC,][]{syred2006review}, which is generated due to a global self-excited instability \cite{Mueller2019}. Although the origin of the PVC is well understood, its impact on combustion performance, especially flame stability, is still a matter of study. The existence of a dominant coherent structure in this flow encourages the idea that manifold learning can be successfully implemented to study the behavior of the PVC under different conditions.

\subsection{Tandem cylinders dataset}\label{subsec:tandem}

The third dataset employed in this work consists of flow field measurements in the wake of tandem cylinders near a wall. The data refer to the work by \citet{RAIOLA2016354}; the experimental configuration is summarized here for completeness. As sketched in figure \ref{fig:tandem_conf_contours}, this configuration consists of two equal cylinders located in a cross-flow which are separated by a ratio of $L/D$ denoted by longitudinal pitch ratio (with $L$ being the longitudinal distance between two cylinders centres and $D$ the diameter of the cylinders equal to $32 mm$). Both cylinders are placed at a similar distance to a wall with a ratio of $W/D$ denoted as wall gap ratio (with $W$ being the distance of the cylinders from the wall). The wind tunnel velocity $U_\infty$ is set constant and equal to $2.3 m/s$ in order to achieve a Reynolds (based on cylinder diameter) number of $4900$ \citep{RAIOLA2016354}. In this study, the longitudinal pitch ratio ($L/D$) is set to $1.5$ and the wall gap ratio ($W/D$) is set to $3$.

While \cite{RAIOLA2016354} reported that a gap ratio $W/D=3$ is sufficient to have negligible wall-interaction effects, three main flow behaviours in the wake of tandem cylinders can be identified based on the Reynolds number and the distances between the two cylinders.
\cite{zdravkovich1997flow} classified the flow around tandem cylinders with identical diameters into three major regimes (extended body, reattachment, and co-shedding), depending on the longitudinal pitch ratio. At low longitudinal distances ($L/D<1.5$), the vortex shedding for the upstream cylinder is suppressed, and the system act as a unified bluff-body, which is categorized as the extended body regime or single bluff-body regime. By increasing the longitudinal distance($1.5<L/D<4$), the flow starts to show some more complex behaviour which mainly can be characterized by the reattachment of the separated free shear layers from the upstream cylinder on the surface of the downstream cylinder. This regime is referred to as `reattachment' regime. Furthermore, by increasing the longitudinal distances, both cylinders start to have their wake with typical characteristics of a Kármán street. This regime is often defined `co-shedding' regime. 
\cite{alam2018vortex} report the existence of a transitional $L/D$ range between the reattachment and co-shedding regimes also referred to as `critical' or `bistable' flow spacing. While it can be argued that a similar bistable regime should occur also between the extended body and the reattachment regimes, \cite{RAIOLA2016354} did not identify such feature for $L/D=1.5$. This dataset appears to be especially suited to discover whether non-linear manifold learning could unveil such a kind of bistable regime.

\section{Results \label{sec:results}}
This section presents and discusses the performance of the proposed encoder-decoder algorithm. The first subsection is dedicated to the performance of the encoder part. We discuss its strengths in unravelling the physical characteristics of the flows distilling the manifolds. In the second subsection, the decoder's ability to reconstruct the original flow fields from the obtained low-dimensional coordinates is analyzed.

\subsection{Encoder's capabilities}
In this section, the embedding manifolds obtained from Isomap are presented for the datasets described in section \ref{sec:data}. A discussion about the choice of the number of neighbours $k$ to build the neighbouring graph $G$ is included in Appendix \ref{appA}. 

In order to compute the dimensionality of the datasets at hand, the residual variance as defined in \eqref{eq:residualvarianceT} is obtained for each number of dimensions. The choice of $p$ is then made based on the elbow method following \cite{Tenenbaum2319}. 
The dimension beyond which the residual variance experiences negligible variation can be identified as the true dimensionality of the dataset. This method is also widely used the compute the proper number of the clusters in the clustering techniques \citep{kaufman2009finding}. The `true' dimensionality is the proper place in a trade-off between the simplicity of the embedded manifold and the loss of information due to truncation. \par

An example of residual variance as a function of the number of selected dimensions is reported in the figure \ref{fig:true_dimensionality} for the case of the wake of the pinball at $Re = 80$ and $Re = 130$. For the lowest Reynolds number, it can be seen that 3 dimensions are sufficient to describe the bulk of the variance since the resulting residual variance for dimensions higher than 3 remains approximately the same. Note that due to the definition of the residual variance, residual variance might be not monotonically decreasing with increasing number of coordinates; for more details the reader is referred to appendix \ref{app:residualvariance}. On the other hand, truncating at three dimensions appears to still be acceptable although induces a larger error in terms of explained variance. 

It must be remarked that, from now on, we are limited in plotting the projection of the manifold on the first 3 dimensions; nonetheless, the number of dimensions needed to represent accurately the manifold might be larger, depending on the complexity of the dynamics. 

\subsubsection{Fluidic pinball}
\begin{figure}
     \centering
        \begin{subfigure}[b]{0.7\textwidth}
        \includegraphics[width=\textwidth]{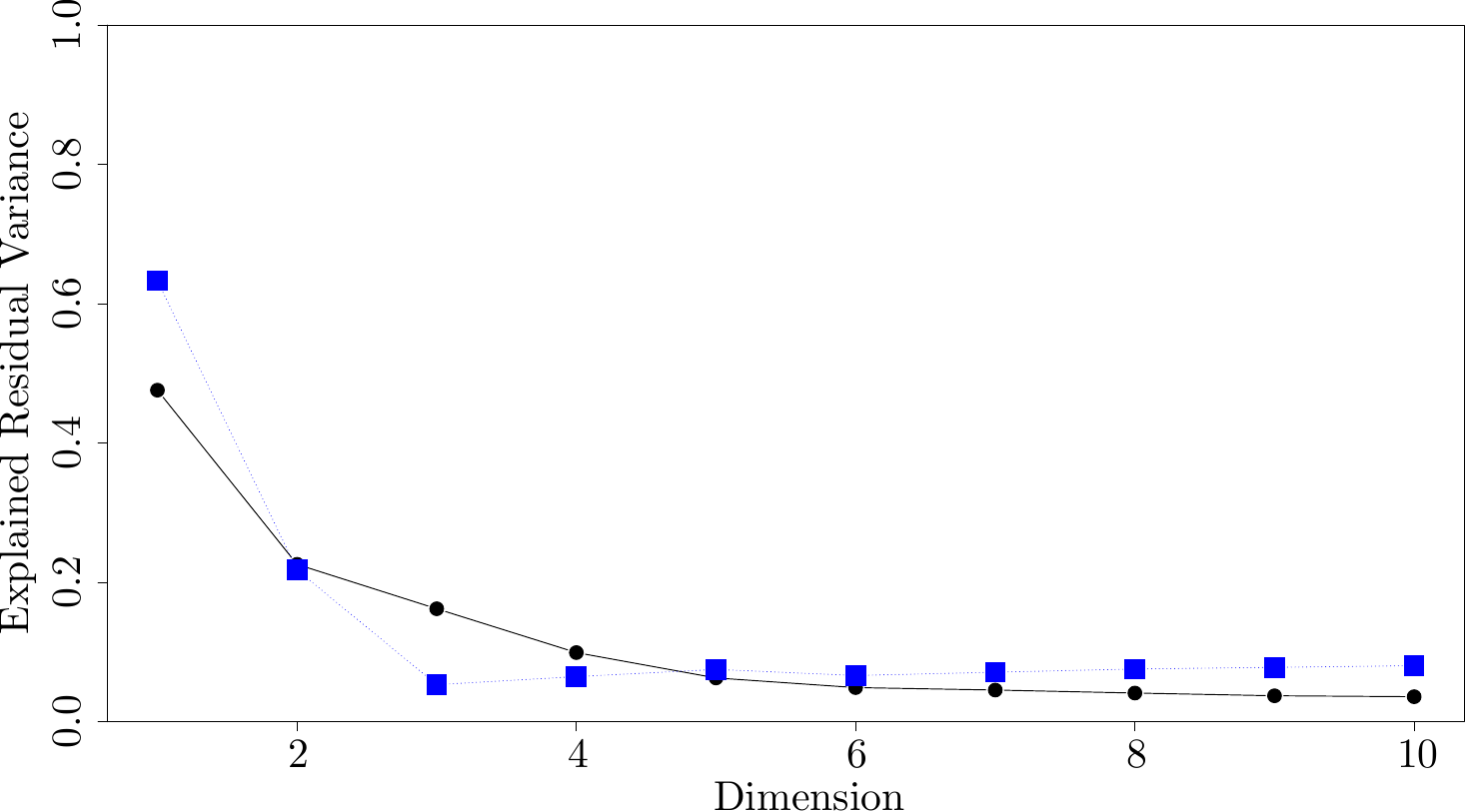}
         \caption{}
        \label{fig:true_dimensionality}
     \end{subfigure}
     \hfill
     \begin{subfigure}[b]{0.32\textwidth}
        \includegraphics[width=\textwidth]{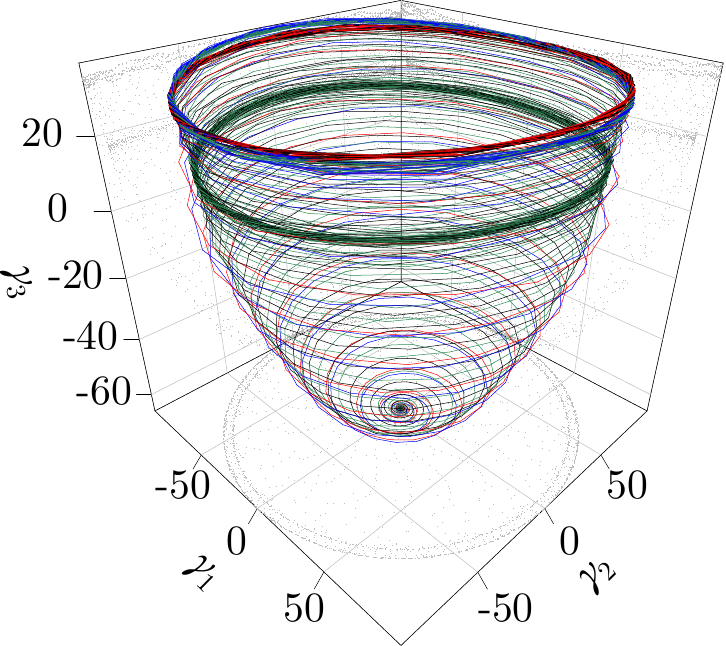}
         \caption{}
        \label{fig:iso_emb_re80_merged_3D_view}
     \end{subfigure}
        \hfill
     \begin{subfigure}[b]{0.32\textwidth}
     \includegraphics[width=\textwidth]{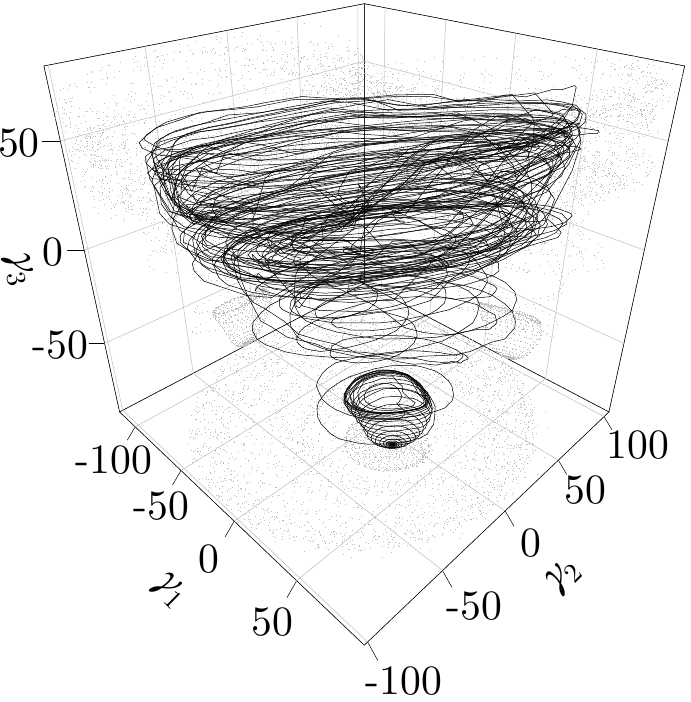}
         \caption{}
     \label{fig:iso_emb_re130_sym_3D_view}
     \end{subfigure}
     \hfill
     \begin{subfigure}[b]{0.32\textwidth}
     \includegraphics[width=\textwidth]{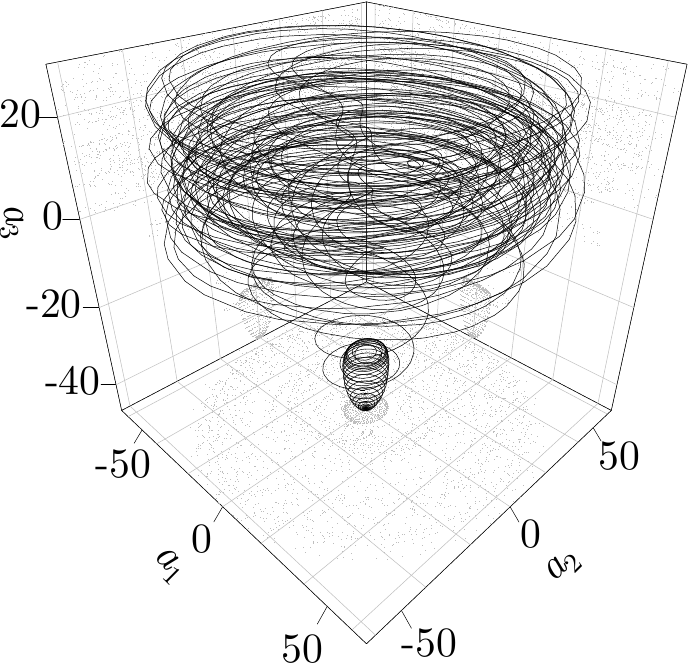}
         \caption{}
     \label{fig:pca_emb_re130_sym_3D_view}
     \end{subfigure}
     \caption{Isomap and POD's results in 3D for the pinball datasets. a) Residual Variance- Blue: $Re = 80$- Black: $Re = 130$- b) Isomap  embedded manifold of $Re = 80$ perspective view.  Black: Symmetric steady solution- Green: Flipped symmetric steady solution- Red: Asymmetric upward steady solution- Blue: Asymmetric downward steady solution. - c) Isomap embedded manifold of $Re = 130$ perspective view- d) POD embedded manifold of $Re = 130$ perspective view.}
     \label{fig:iso_emb_re80_merged}
\end{figure}

Figure \ref{fig:true_dimensionality} shows the residual variances of the pinball configuration dataset for $Re = 80$ and $p=1,\ldots,10$. For this configuration, the chosen value of $k$ is equal to 8; for more details about the choice of $k$ the reader is referred to appendix \ref{appA}
The elbow is attained for $p=3$ which is considered to be the true dimensionality of this problem.
The residual variance is approximately the same for $p>3$, thus it can be argued that the manifold of input data, embedded in a higher dimensional state space, has three key dimensions. 

The same procedure has been done for $Re = 130$ case, employing $k=12$.
The residual variance decreases monotonically for increasing number of dimensions and is already below 20\%  after three-dimensions ($p=3$) (black curve in figure \ref{fig:true_dimensionality}). 
However, the residual variance value at $p=3$ increases by increasing the Reynolds number to higher values and entering the more chaotic regimes.
This indicates that in the chaotic regime the 'true' dimensionality is higher due to the arising of a more complex dynamics. Therefore, it is reasonable to expect that, for increasing Reynolds number, the number of dimensions needed to explain the bulk of the variance should increase. This behaviour is not surprising, and it is observed in virtually all dimensionality reduction techniques. 

In terms of manifold shape, both for $Re = 80$ and $130$ data lies on a paraboloid with the first two coordinates ($\gamma_1$ and $\gamma_2$) being representative of the periodic vortex shedding and the third coordinate being representative of a shift-mode characteristic of the transient dynamics from the onset of vortex shedding to the periodic von Kármán wake, analogously to what found for the cylinder wake by \cite{noack2003hierarchy}. A similar paraboloid shape is reported for the fluidic pinball at $Re=30$ by \cite{Deng_2019}. It is remarkable that, when analyzing all the solutions at $Re=80$, the manifold correctly identifies the first unstable limit cycle for the symmetric unstable solution and is able to identify the differences between the asymmetric upward and downward limit cycle. As well for $Re=130$, the chaotic nature of the data shows a less smooth manifold, which is still possible to visualize the characteristic paraboloid shape.
When employing POD and plotting the first three temporal modes, similar shapes could be obtained although less clear, expecially at $Re=130$ as evident from the comparison of figures \ref{fig:iso_emb_re80_merged} (c) and (d).

\begin{figure}
     \centering
     \begin{subfigure}[b]{0.49\textwidth}
        \centering
        \includegraphics[width=\textwidth]{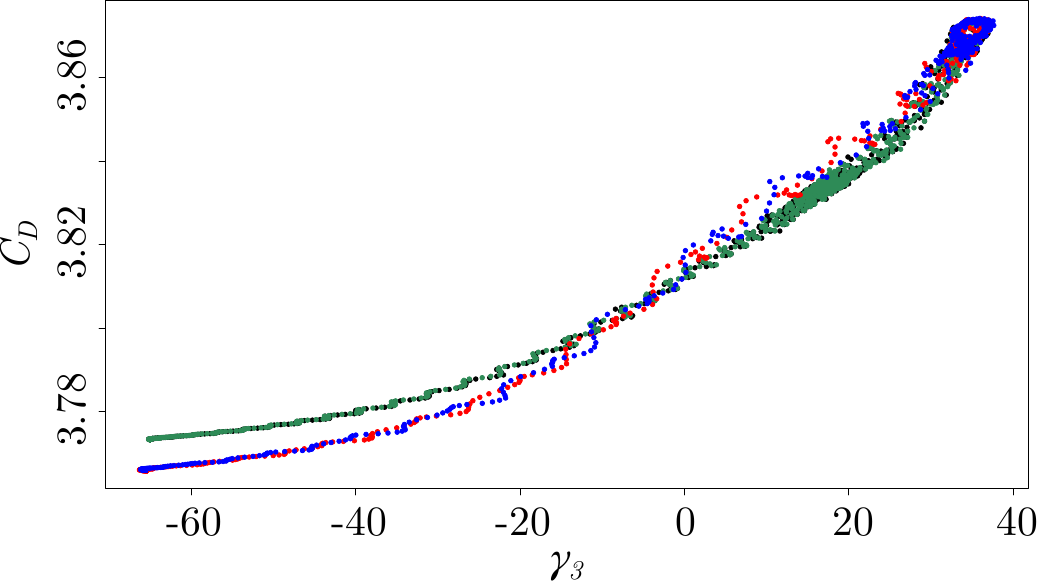}
         \caption{}
        \label{fig:cd_vs_dim3}
     \end{subfigure}
     \hfill
     \begin{subfigure}[b]{0.49\textwidth}
         \centering
        \includegraphics[width=\textwidth]{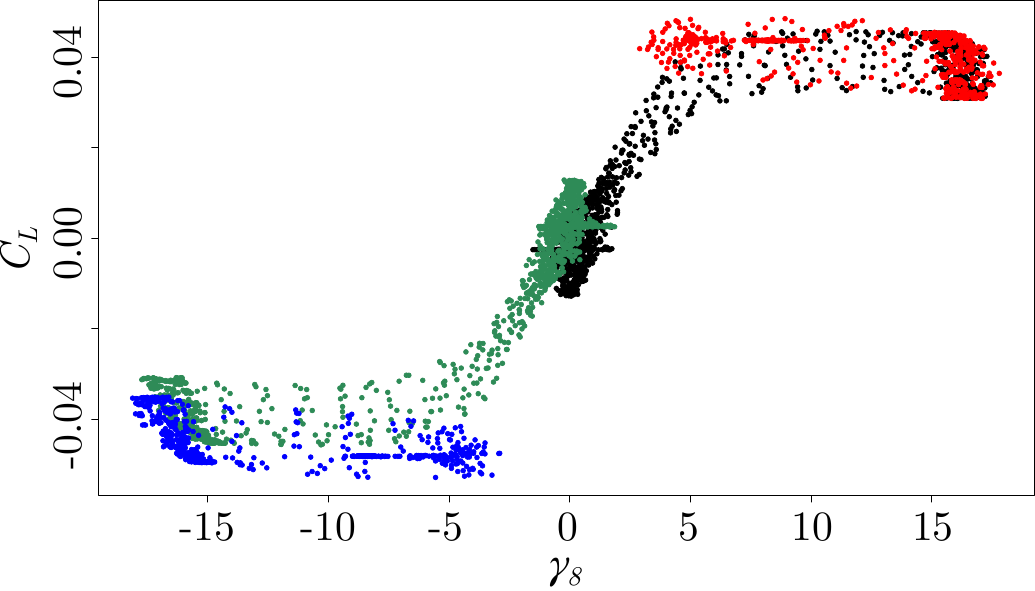}
         \caption{}
        \label{fig:cl_vs_dim6}
     \end{subfigure}
        \caption{Relation between the coordinates of the embedded manifold and the force coefficients for the fluidic pinball at $Re=80$: a) the drag coefficient ($C_D$) versus $3rd$ coordinate, b) the lift coefficient ($C_L$) versus $8th$ coordinate. 
        Black: Symmetric steady solution- Green: Flipped symmetric steady solution- Red: Asymmetric upward steady solution- Blue: Asymmetric downward steady solution.}
        \label{fig:force_relation}
\end{figure}

Although the dimensions  identified by Isomap do not necessarily have a physical meaning, it is a useful exercise to establish whether there exists some correlation between such coordinates and relevant flow quantities.
Figure \ref{fig:force_relation}(a) shows a clear correlation between the drag coefficient and the coordinate $\gamma_3$. This correlation is reasonably expected, since we observed that the third coordinate is representative of the shift mode. Some pairs of coordinates are related to higher-order harmonics of the flow, as is the case for $\gamma_1 - \gamma_2$, $\gamma_4 - \gamma_5$, and $\gamma_6 - \gamma_7$. 
Interestingly enough, if we extend the analysis to higher-order coordinates, it is possible to identify a high degree of correlation between the $\gamma_8$ and the lift coefficient $C_L$, as shown in figure \ref{fig:force_relation}(b).
Although these interpretations are case-sensitive and can be affected by changes in the flow configuration and starting conditions in the Isomap algorithm, we have identified situations in which some coordinates in the low-dimensional space could be related to the main flow features. While it is outside of the scope to assess the interpretability of the coordinates identified by Isomap, we spotlight the possibility of existence of such kind of correlations. This could be a powerful catalyst for the extension of the encoder-decoder framework presented here, and it will be object of future study.  

\begin{figure}
     \centering
     \begin{subfigure}[b]{0.7\textwidth}
         \centering
        \includegraphics[width=\textwidth]{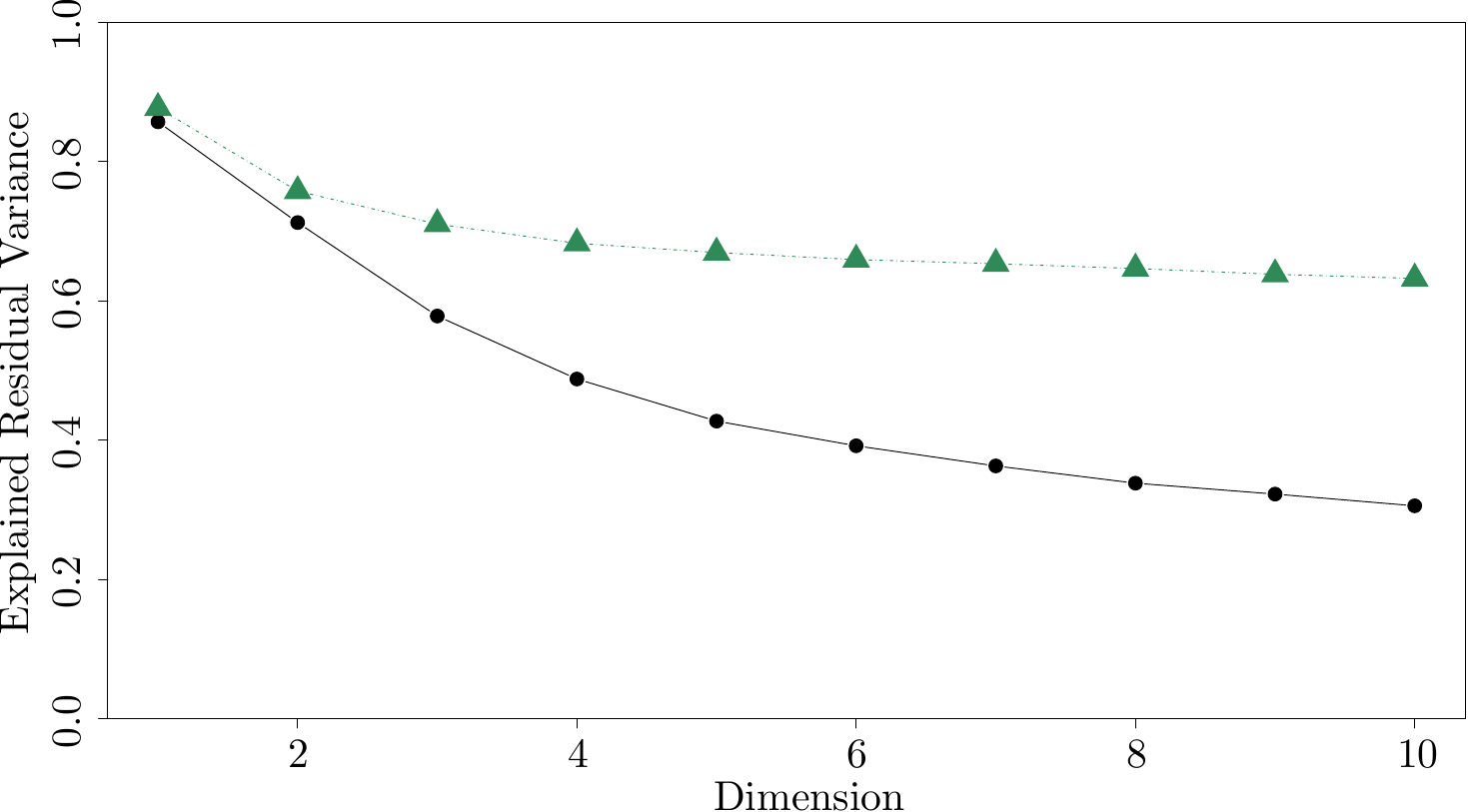}
         \caption{}
        \label{fig:res_var_swj}
     \end{subfigure}
     \hfill
     \begin{subfigure}[b]{0.49\textwidth}
        \includegraphics[width=\textwidth]{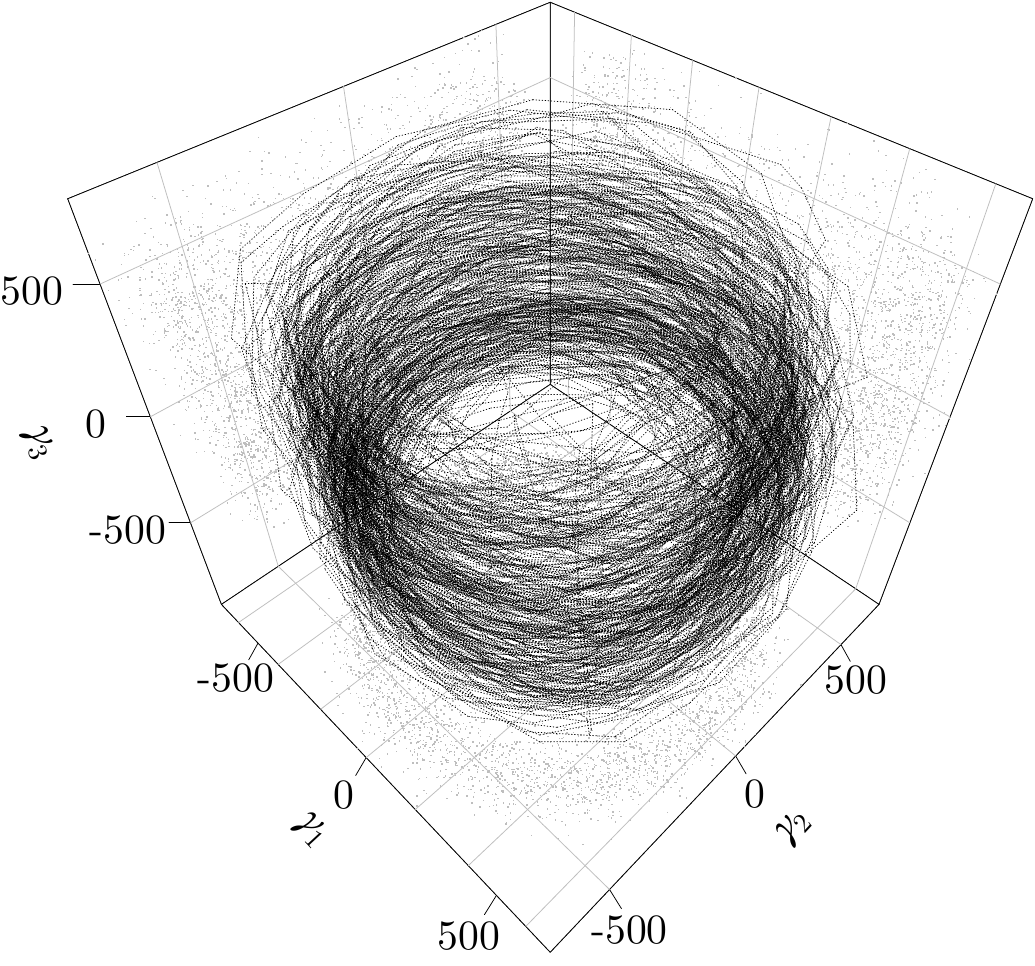}
         \caption{}
        \label{fig:Isomap_emb_3d_swj}
     \end{subfigure}
     \hfill
     \begin{subfigure}[b]{0.49\textwidth}
        \includegraphics[width=\textwidth]{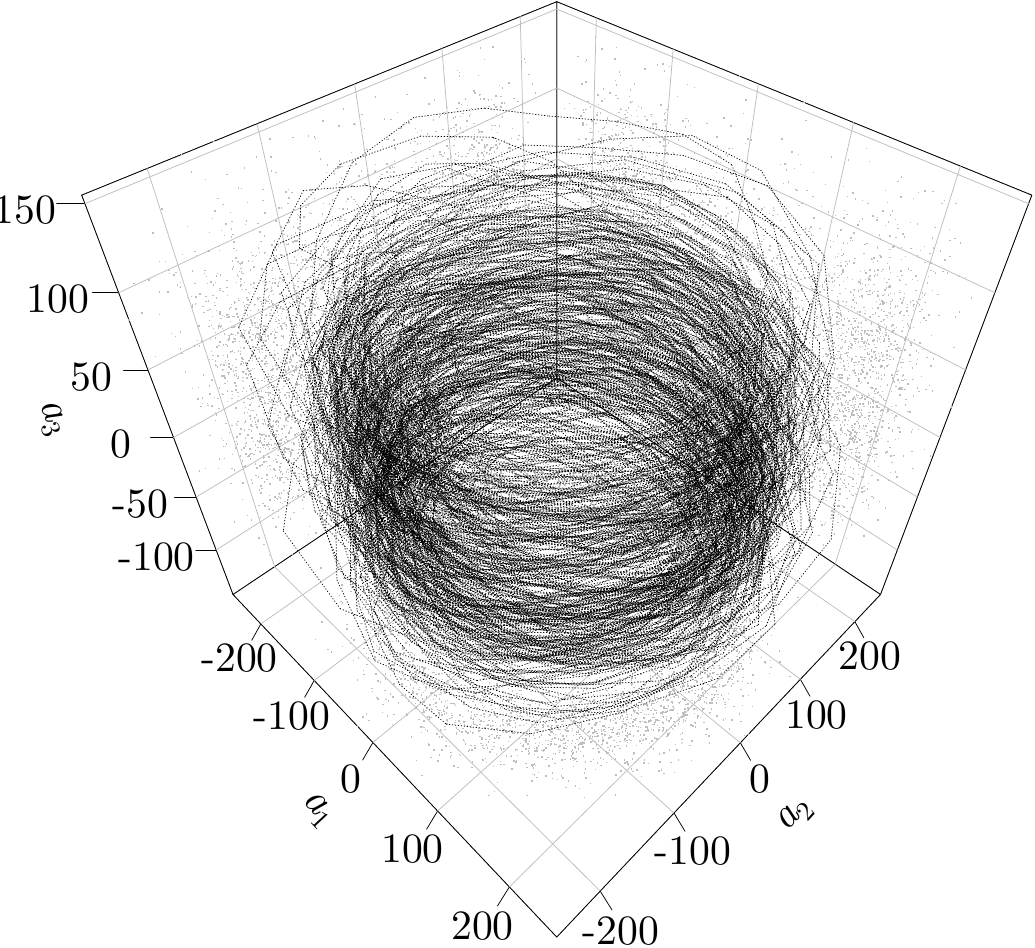}
         \caption{}
        \label{fig:POD_emb_3d_swj}
     \end{subfigure}
     \hfill
     \begin{subfigure}[b]{0.49\textwidth}
         \centering
        \includegraphics[width=\textwidth]{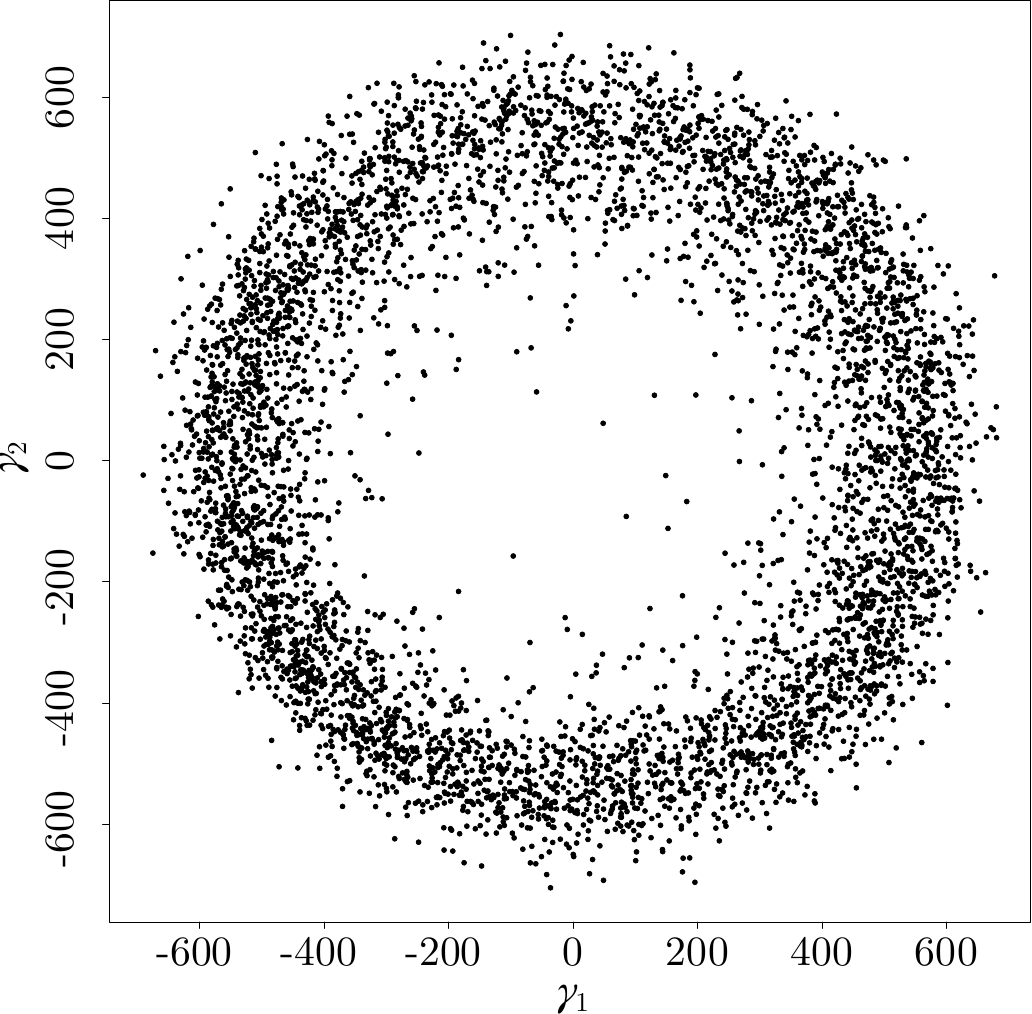}
         \caption{}
        \label{fig:Isomap_emb_top_swj}
     \end{subfigure}
     \hfill
     \begin{subfigure}[b]{0.49\textwidth}
         \centering
        \includegraphics[width=\textwidth]{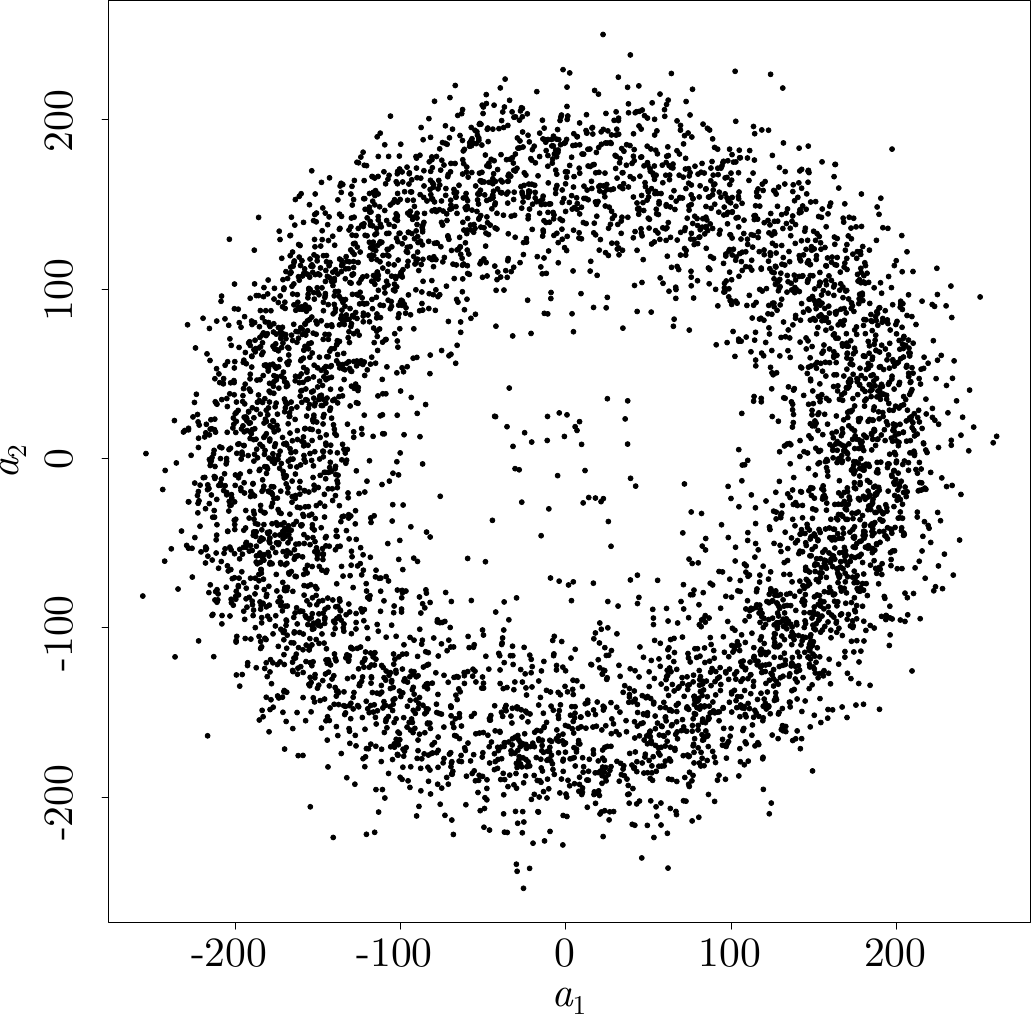}
         \caption{}
        \label{fig:POD_emb_top_swj}
     \end{subfigure}
        \caption{Resulted manifold in 3D space from Isomap and POD for swirling jet case. a) Residual variance; Green: POD, Black: Isomap- b) 3D Isomap- c) 3D POD- d) Top view- Isomap- e) Top view- POD.}
        \label{fig:iso_POD_top_embedding_swj}
\end{figure}

\subsubsection{Swirling jet}
In the more challenging experimental case of the swirling jet, the same encoder procedure based on Isomap using $k = 8$ has been carried out. POD and Isomap performance as encoders is compared. 

Figure \ref{fig:iso_POD_top_embedding_swj}(a) shows the residual variances of Isomap (black dots) and  POD (green triangles), which is measured as stated in Appendix \ref{app:residualvariance}, for different numbers of dimensions. In this case, the values of residual variance are significantly larger compared to those in the simulation cases studied before most likely due to the turbulent nature of the flow and possibly due to measurement noise in the experimental data. 
It is worth noting that the residual variances of Isomap are lower than those of POD for all the dimensions depicted, thus indicating that Isomap is a better manifold learner in this case to preserve the geometry of high-dimensional dataset. To investigate this advantage, we can compare the resulting embedding in both cases for three dimensions, which accounts for a reasonable amount of residual variance.
Figures \ref{fig:iso_POD_top_embedding_swj}(b) and \ref{fig:iso_POD_top_embedding_swj}(d) show the encoded data by Isomap, whereas figures \ref{fig:iso_POD_top_embedding_swj}(c) and \ref{fig:iso_POD_top_embedding_swj}(e) illustrate the POD results, respectively.
Although for both methods the general shape of the embedded manifold is similar to a hollow cylinder, the one obtained by Isomap shows a more clearly defined shape. Furthermore, the diameter of the hollow cylinder in POD is smaller  and less circular with more spread points. Thus, the Isomap encoding  provides a more helpful base to interpret the manifold and eventually relate the low dimensional coordinates to the physical features of the flow. 

\subsubsection{Tandem cylinders}
\begin{figure}
     \centering
     \begin{subfigure}[b]{0.7\textwidth}
         \centering
        \includegraphics[width=\textwidth]{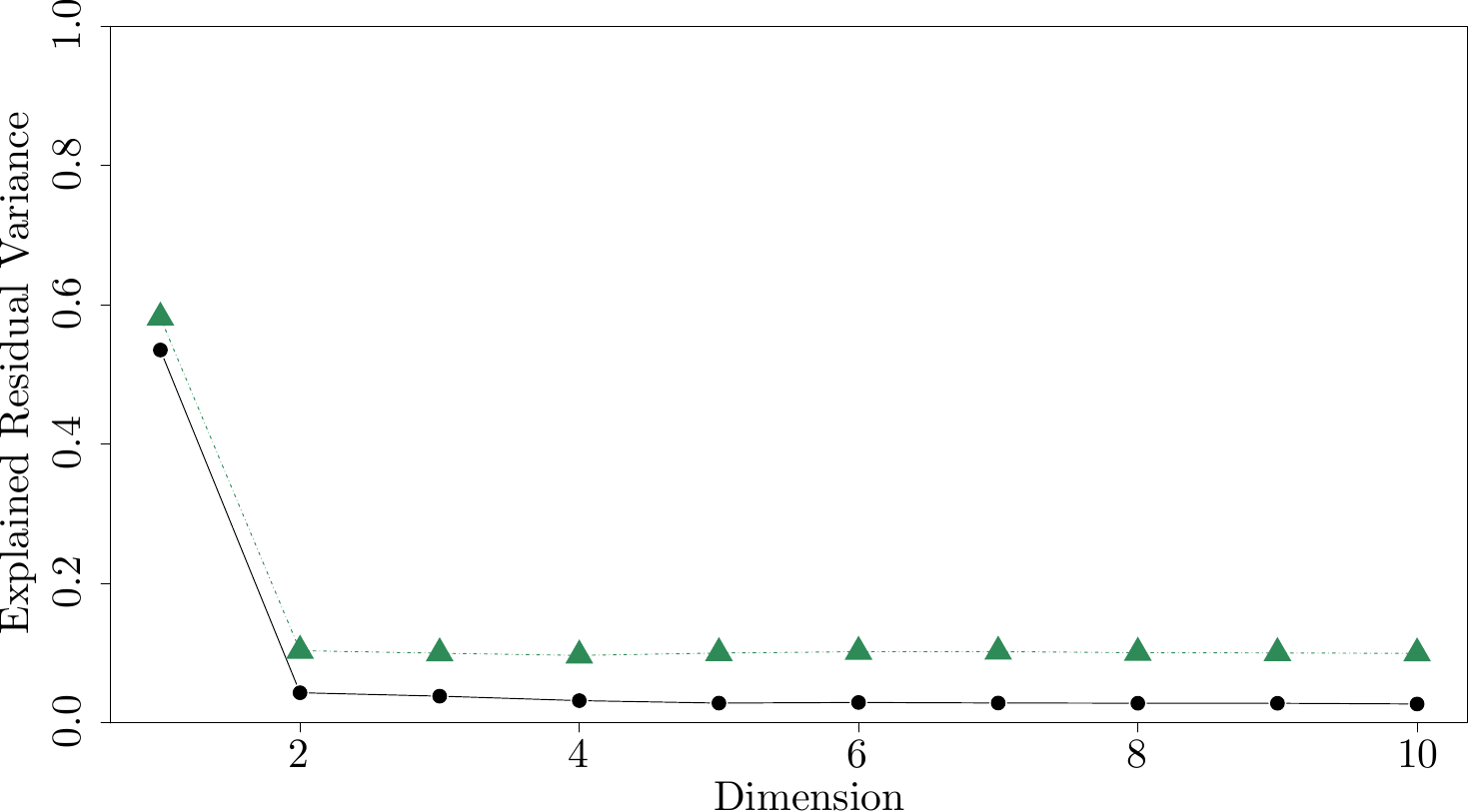}
         \caption{}
        \label{fig:res_var_tandem}
     \end{subfigure}
     \hfill
     \begin{subfigure}[b]{0.49\textwidth}
         \centering
        \includegraphics[width=\textwidth]{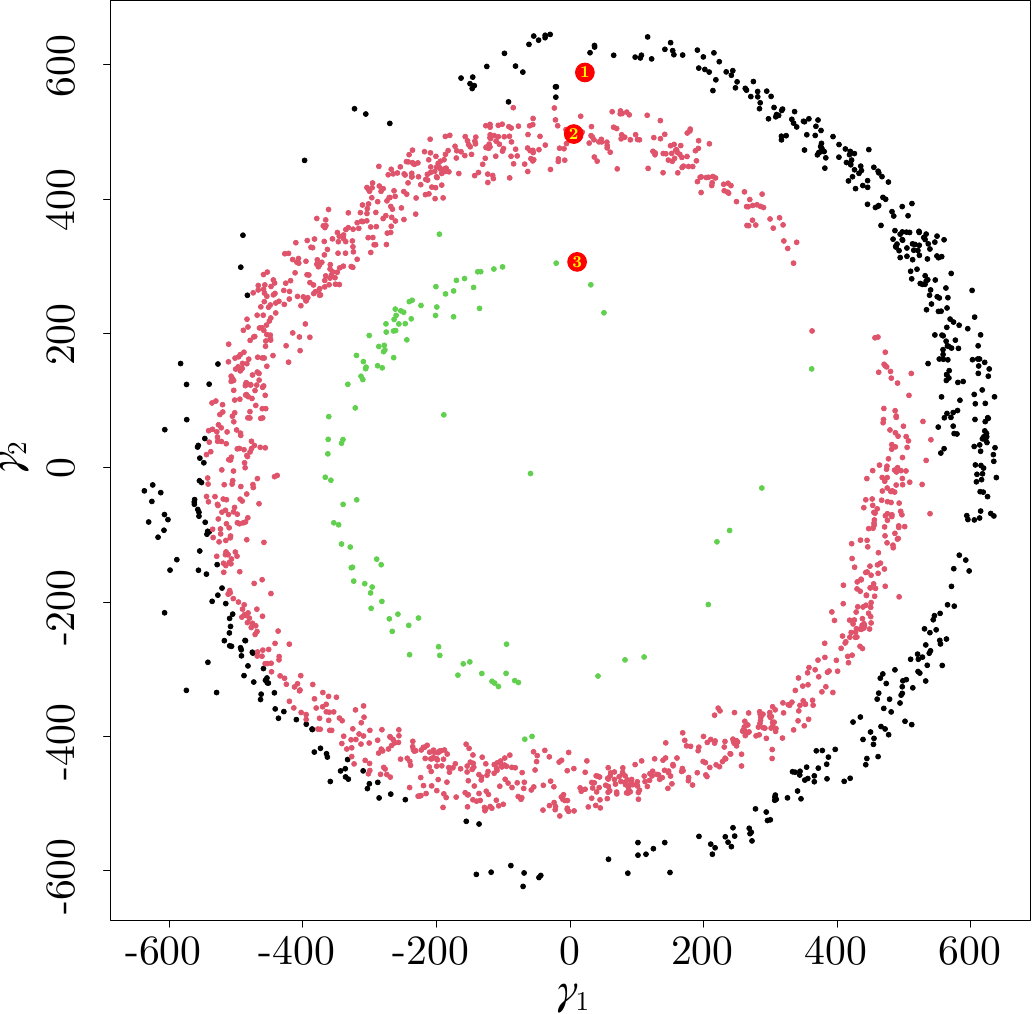}
         \caption{}
        \label{fig:Isomap_emb_top_tandem}
     \end{subfigure}
     \hfill
     \begin{subfigure}[b]{0.49\textwidth}
         \centering
        \includegraphics[width=\textwidth]{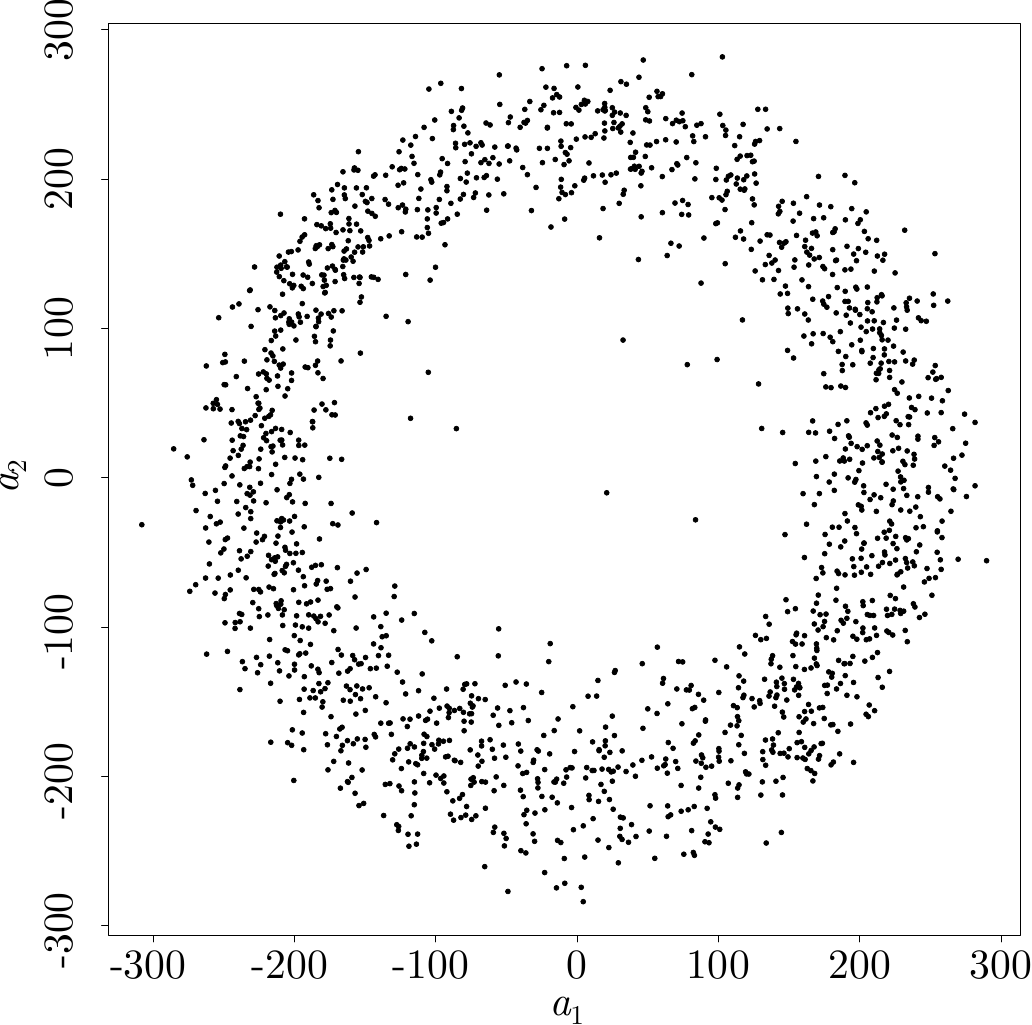}
         \caption{}
        \label{fig:POD_emb_top_tandem}
     \end{subfigure}
     \hfill
     \begin{subfigure}[b]{0.32\textwidth}
         \centering
        \includegraphics[width=\textwidth]{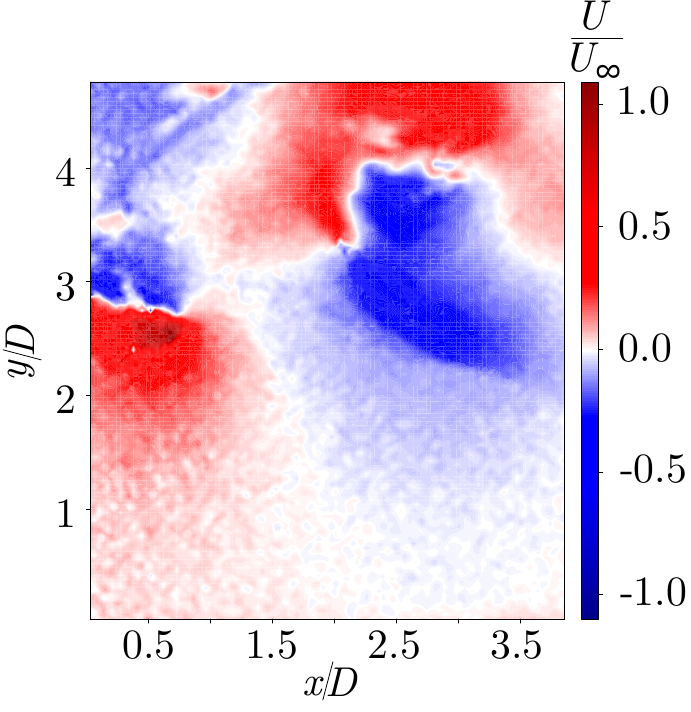}
         \caption{}
        \label{fig:tandem_sample4}
     \end{subfigure}
          \hfill
     \begin{subfigure}[b]{0.32\textwidth}
         \centering
        \includegraphics[width=\textwidth]{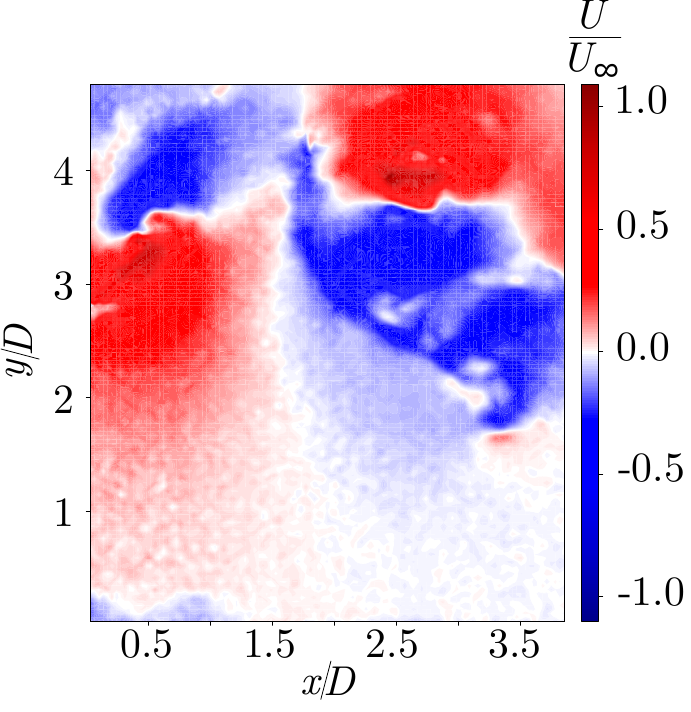}
         \caption{}
        \label{fig:tandem_sample3}
     \end{subfigure}
          \hfill
       \begin{subfigure}[b]{0.32\textwidth}
         \centering
        \includegraphics[width=\textwidth]{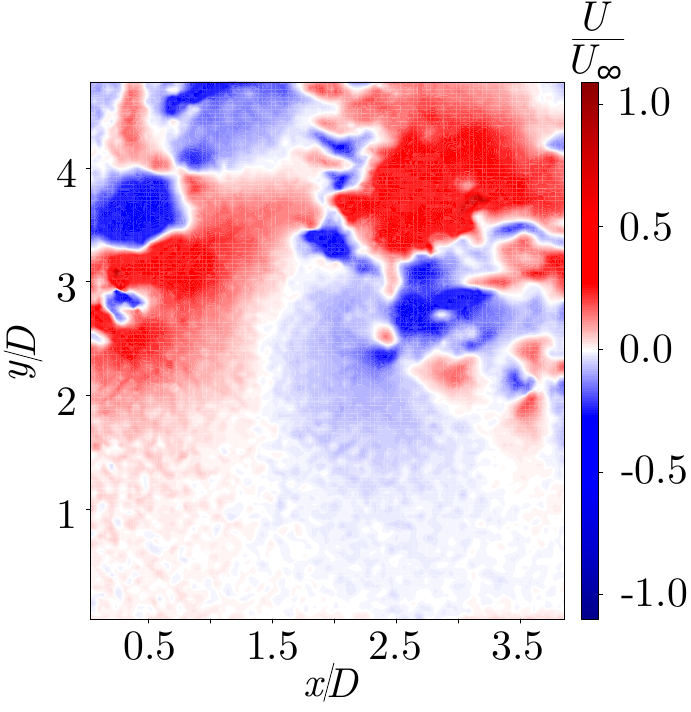}
         \caption{}
        \label{fig:tandem_sample1}
     \end{subfigure}
        \caption{Resulted manifold in 2D space from Isomap and POD for the tandem cylinders case. a) Residual variance; Green: POD, Black: Isomap- b) Isomap grouped embedding manifold; Black: Group 1- Red: Group 2- Green: Group 3- c) POD.- d to f) Sample snapshots for each group.}
        \label{fig:iso_POD_top_embedding_tandem}
\end{figure}

Regarding the tandem cylinder dataset, Figure \ref{fig:iso_POD_top_embedding_tandem}(a) shows that an appropriate number of dimensions in terms of the residual variance for both Isomap (black dots) and POD (green triangles) is two.
Furthermore, as in the previous case, Isomap residual variances outperforms those of POD.
As in the swirling-jet case discussed above, the Isomap results in figure \ref{fig:iso_POD_top_embedding_tandem}(b) show a clearer manifold compared to the one obtained by POD (figure \ref{fig:iso_POD_top_embedding_tandem}(c)). Furthermore, it also shows some separate groups of snapshots related to some physical features of the flow while the manifold resulted from POD ultimately fails to capture this behavior of the system. 
Setting the number of groups to three, the results of the classification in the polar coordinates are shown by means of different colors in figure \ref{fig:iso_POD_top_embedding_tandem}(b) 
The flow fields of each group close to $\gamma_1=0$ line are plotted in figures \ref{fig:iso_POD_top_embedding_tandem}(d)-(f) from the outer group to the inner one, as the prototype of each group.
Investigating each group by looking at these prototypes, which are highlighted as red dots in figure \ref{fig:iso_POD_top_embedding_tandem}(b), reveals that as we move outward from the center ($\gamma_1=\gamma_2=0$), the distance between two consecutive vortices in the wake decreases, and the vortices appear less intense. In other words, moving from the groups from the outer part to the inner one it is found a transition between two different vortex shedding regimes which could correspond to the bluff-body regime and the reattachment regime of tandem cylinders. As described in \S \ref{subsec:tandem}, previous studies suggest that this configuration of the flow lies on the bluff-body regime, as most of the snapshots in low-dimensional space classify in the outer and middle groups, and the results of our encoding procedure are consistent with previous studies done by using POD \citep{RAIOLA2016354}. However, Isomap can capture that even in this configuration, some behavior of the next regime can coexist with the dominant one, suggesting that we cannot define a specific number for $L/D$ as the classifier of the flow regimes, and by increasing $L/D$ the flow smoothly changes its behaviors.  

\subsection{Decoder's performance}

This section assesses the quality of the decoder approach described in \S \ref{sec:dim_red_methods}. The normalized mean squared error (NMSE) in a test set of observations $\mathcal{T}$ is computed as

\begin{equation}\label{eq:mse}
 NMSE = \frac{1}{|\mathcal{T}|}\sum_{\mathbf{x}\in\mathcal{T}} \frac{(\mathbf{x} - \hat{\mathbf{x}})^2}{\|\mathbf{x}\|},
\end{equation}

where $\hat{\mathbf{x}}$ is the decoder's reconstruction of the flow field $\mathbf{x}\in\mathbb{R}^P$.

\begin{figure}
     \centering
     \begin{subfigure}[b]{0.49\textwidth}
         \centering
        \includegraphics[width=\textwidth, interpolate]{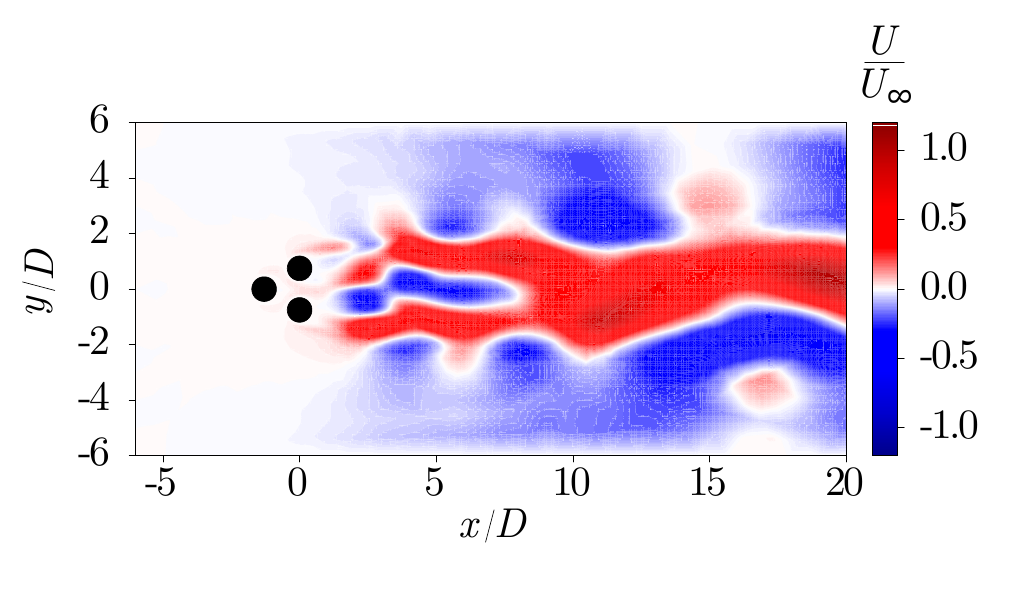}
         \caption{}
        \label{fig:decoder_recons_Re130}
     \end{subfigure}
     \hfill
     \begin{subfigure}[b]{0.49\textwidth}
         \centering
        \includegraphics[width=\textwidth]{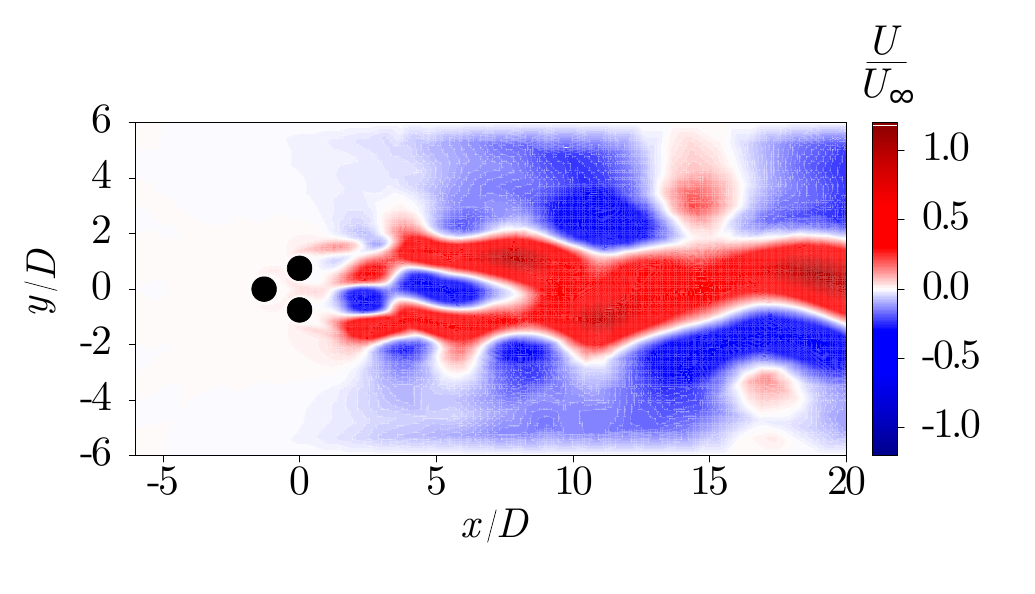}
         \caption{}
        \label{fig:decoder_actual_Re130}
     \end{subfigure}
          \hfill
     \begin{subfigure}[b]{0.49\textwidth}
         \centering
        \includegraphics[width=\textwidth]{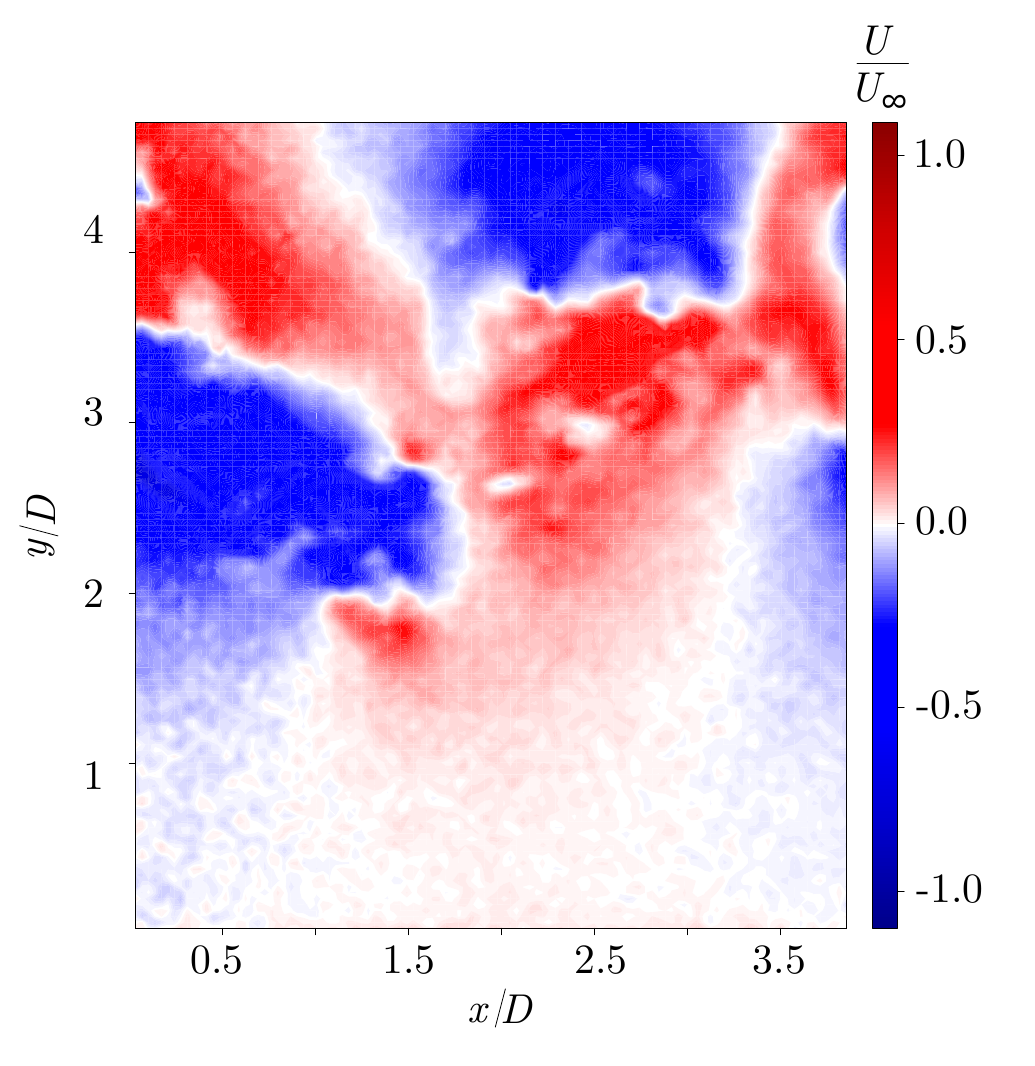}
         \caption{}
        \label{fig:decoder_recons_Tandem}
     \end{subfigure}
          \hfill
     \begin{subfigure}[b]{0.49\textwidth}
         \centering
        \includegraphics[width=\textwidth]{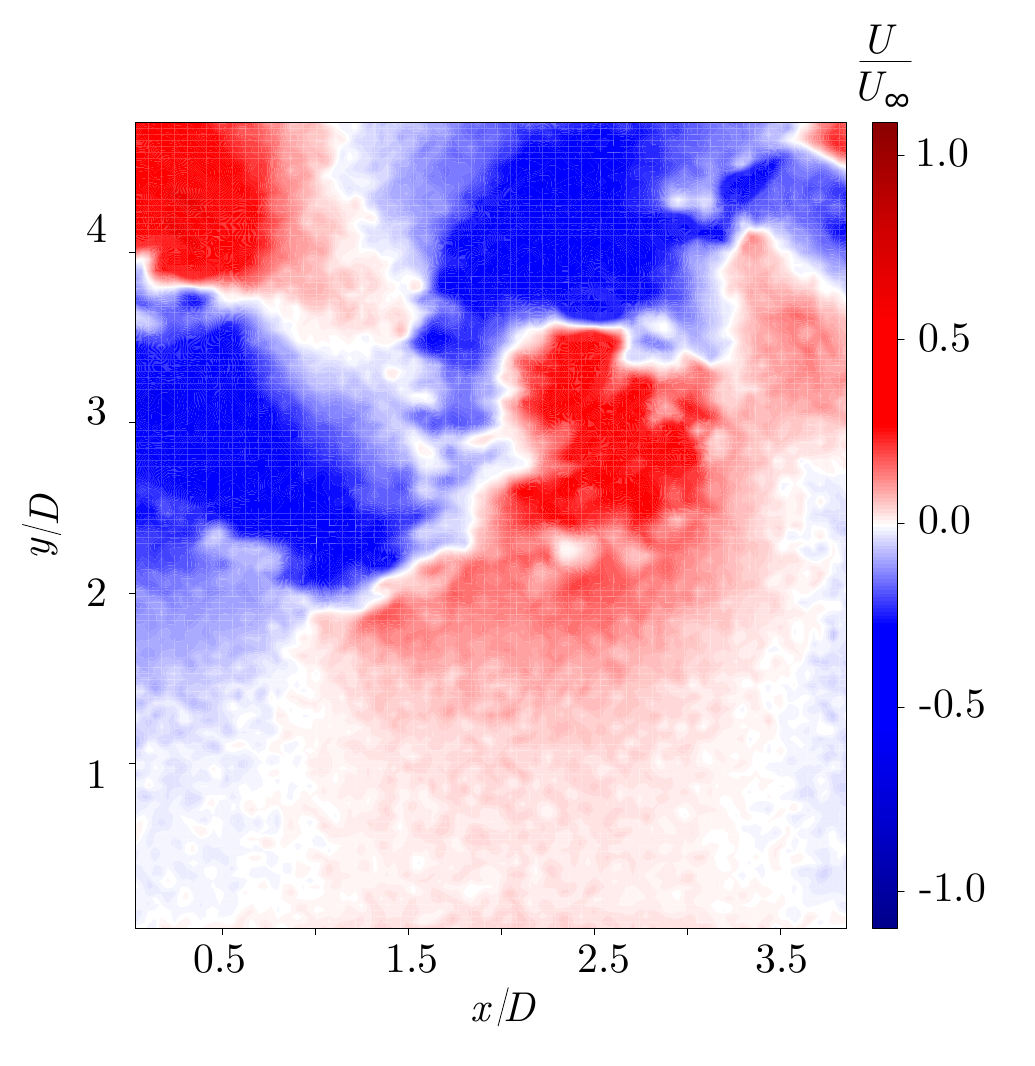}
         \caption{}
        \label{fig:decoder_actual_Tandem}
     \end{subfigure}
        \caption{Reconstructed flow field using $K$NN decoder on the embedded manifold of the Isomap. a) Reconstructed- Pinball $Re = 130$- b) Actual- Pinball $Re = 130$- c) Reconstructed- Tandem cylinders- d) Actual- Tandem cylinders.}
        \label{fig:decoder_performance}
\end{figure}

Table \ref{tab:kd} shows the average NMSE obtained (last column) for the different cases studied, namely Pinball for $Re = 80$ and $Re = 130$, the swirling jet flow and the wake of tandem cylinders. The sample data has been split into $70\%$ training and $30\%$ testing. The number of neighbors considered in Isomap ($k$) and the number of neighbors considered in the decoding stage ($K$) are reported in the sixth and seventh columns, respectively. The first column in Table \ref{tab:kd} depicts the case study, the second the value of $Re$, the third the number of samples, fourth the number of features in the high-dimensional space and fifth the the number of dimensions considered for Isomap. Observe that the NMSEs reported in Table \ref{tab:kd} are actually representation errors, i.e.\ the error incurred when reconstructing the original datasets by means of our encoder-decoder methodology. These representation errors show that the decoder has an excellent performance on the simulation datasets and has a reasonably good performance on both noisy experimental datasets. Figure \ref{fig:decoder_performance} compares reconstructed and actual snapshots for a random point in the test subset. As expected from the reported errors, the differences between the two snapshots in the pinball case are hardly noticeable. In the case of the tandem cylinders, although there are some noticeable differences between reconstructed and actual snapshots, the general behavior of the flow is well preserved in the reconstructed snapshot.

\begin{table}
  \begin{center}
\def~{\hphantom{0}}
  \begin{tabular}{lccccccc}
      Case  & $Re$ & $N$ & $P$ & $p$  &   $k$ & $K$ & Average NMSE  \\[3pt]
       Pinball & 80 & 4000 & 155358 & 3 & 8 & 6 & 2.10 \\
       Pinball & 130 & 4000 & 155358 & 3 & 12 & 6 & 7.25 \\
       Tandem Cylinders & 4900 & 1800 & 30414 & 2 & 3 & 4 & 37.50 \\
       Swirling Jet & 20000 & 5396 & 26004 & 3 & 8 & 6 & 41.05 \\
  \end{tabular}
  \caption{Manifold representation errors by using 70 percent of the dataset as training dataset.}
  \label{tab:kd}
  \end{center}
\end{table}

A comparison using as input low-dimensional coordinate resulting from Isomap and POD has been carried out to investigate further the decoder performance. The NMSE of the Isomap decoder was calculated using the same $70\%$ training dataset. The reconstructon error with POD modes was calculated as the reconstruction error employing the first two or three modes, depending on the true dimensionality $p$ of the dataset. Results show a clear superiority of the $K$NN decoder in all the cases except for the tandem-cylinders one. In this case, the performance of the two methods is similar, probably due to the rather small size of the input experimental dataset.\par

\begin{figure}
  \centerline{\includegraphics[width=0.5\textwidth]{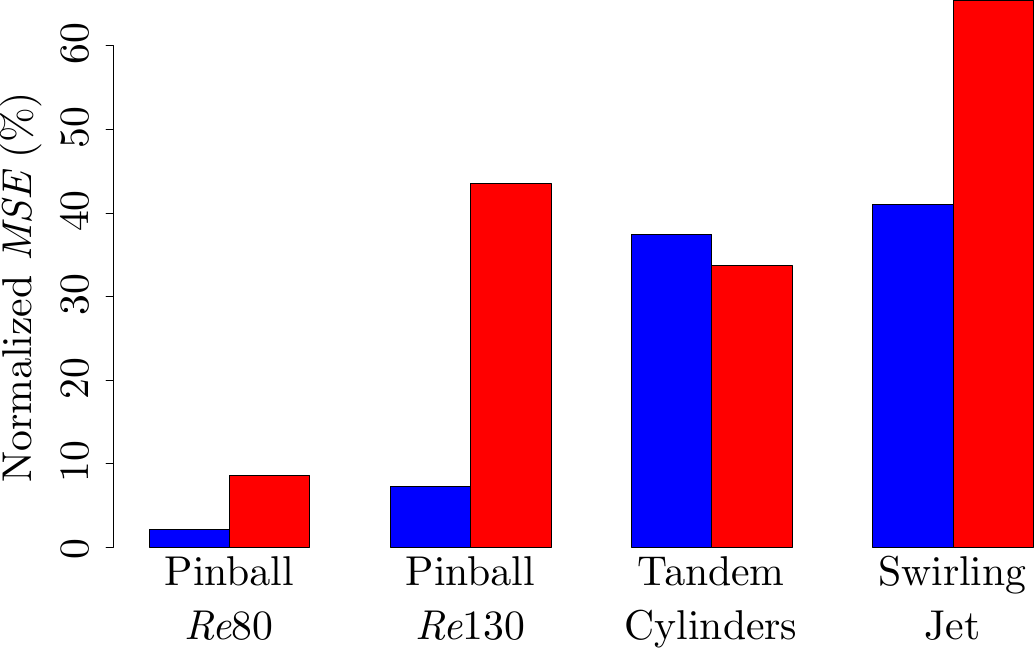}}
  \caption{Reconstruction Error of Isomap vs POD. Blue: Isomap, Red: POD.}
\label{fig:decoder_errors_comparison}
\end{figure}

\section{Conclusions}
\label{sec:conclusion}
In this paper we have developed an encoder-decoder framework based on manifold learning techniques to tackle the problem of understanding shear flows. 
The proposed manifold learner is Isomap and it is coupled with a $K$NN decoder. We show that flows which can be described with a limited set of coherent structures are suitable candidates for manifold learning. In the applications proposed in this manuscript we have chosen phenomena whose snapshots correctly sampled all the events defining the system ``clock'', i.e.\ jet and wake flows in which the measurement domain samples the evolution of the main vortical features.
We have shown that the manifolds unraveled using the Isomap encoder are representative of meaningful physical quantities and are suitable for a reduced-order modelling of shear flows. 
The pure physics-uninformed results in the fluidic pinball case also have a correlation with physical properties like vortex-shedding phases or the force coefficients and open ground to use this technique to model wake flows and design flow-control systems.

We have shown that when handling experimental cases with complex behavior, despite the presence of acceptable measurement noise, the new encoder-decoder tool outperforms the classical dimensionality-reduction techniques like POD in terms of clarity and interpretability of the identified manifolds, and it is less sensitive to experimental noise. In such cases, not only the identified manifolds are more reliable, but they also distill some physical information that POD is not able to catch, including the coexistence of the two shedding regimes and transition between them in the case of the wake of two tandem cylinders. 

Finally, the developed decoder proved to have outstanding capabilities in reconstructing the original flow fields from the identified manifold, allowing for the instantaneous identification of the flow state with applications to closed-loop flow control.

The proposed  manifold learner may significantly reduce the dimension of the state space
as compared to POD expansions with similar representation error,
particularly for transient flows.
Hence, the autoencoder methodology 
may be used for full-state estimation and control design.
For both tasks, every unnecessary coordinate
acts as a noise amplifier and constitutes a danger for the system to get `off track'.
The authors actively pursue this direction.

\section*{Acknowledgements} This work has been supported by the Madrid Government (Comunidad de Madrid) under the Multiannual Agreement with Universidad Carlos III de Madrid in the line of ``Fostering Young Doctors Research" (PITUFLOW-CM-UC3M), and in the context of the V PRICIT (Regional Programme of Research and Technological Innovation).
This work has also been partially supported by the project ARTURO, ref. PID2019-109717RBI00/AEI/10.13039/501100011033, funded by the Spanish State Research Agency.
BRN acknowledges funding 
by the National Science Foundation of China (NSFC) through grants 12172109 and 12172111,
by a  Natural Science and Engineering grant of Guangdong province, China,
and by scientific support from the HangHua company (Dalian, China).

\section*{Declaration of interests} The authors report no conflict of interest.

\section*{Data availability statement} The data that support the findings of this study are available from the corresponding author upon reasonable request.

\appendix
\section{Selection of the number of neighbors $k$ in Isomap \label{appA}}

As discussed in section \ref{sec:dim_red_methods}, selecting a proper number of neighbors $k$ to construct the neighbouring graph $G$ in Isomap is a crucial decision. The unknown geodesic distances between every pair of observations in the high-dimensional manifold are approximated by the shortest pats distances between the corresponding nodes in the graph $G,$ which depends on $k.$
On the one hand, a too small $k$ may cause a splitting of the manifold into disjoint ones and thus losing  its real structure. 
On the other hand, if $k$ is too large then points which are far according to the  real geodesic distance  may become close by using its approximation by  their shortest path  in $G$ due to an undue number of connections (edges) and/or the existence of holes in the manifold. 
This phenomenon is  known as short-circuiting. 

\begin{figure}
  \centerline{\includegraphics[width=0.7\textwidth]{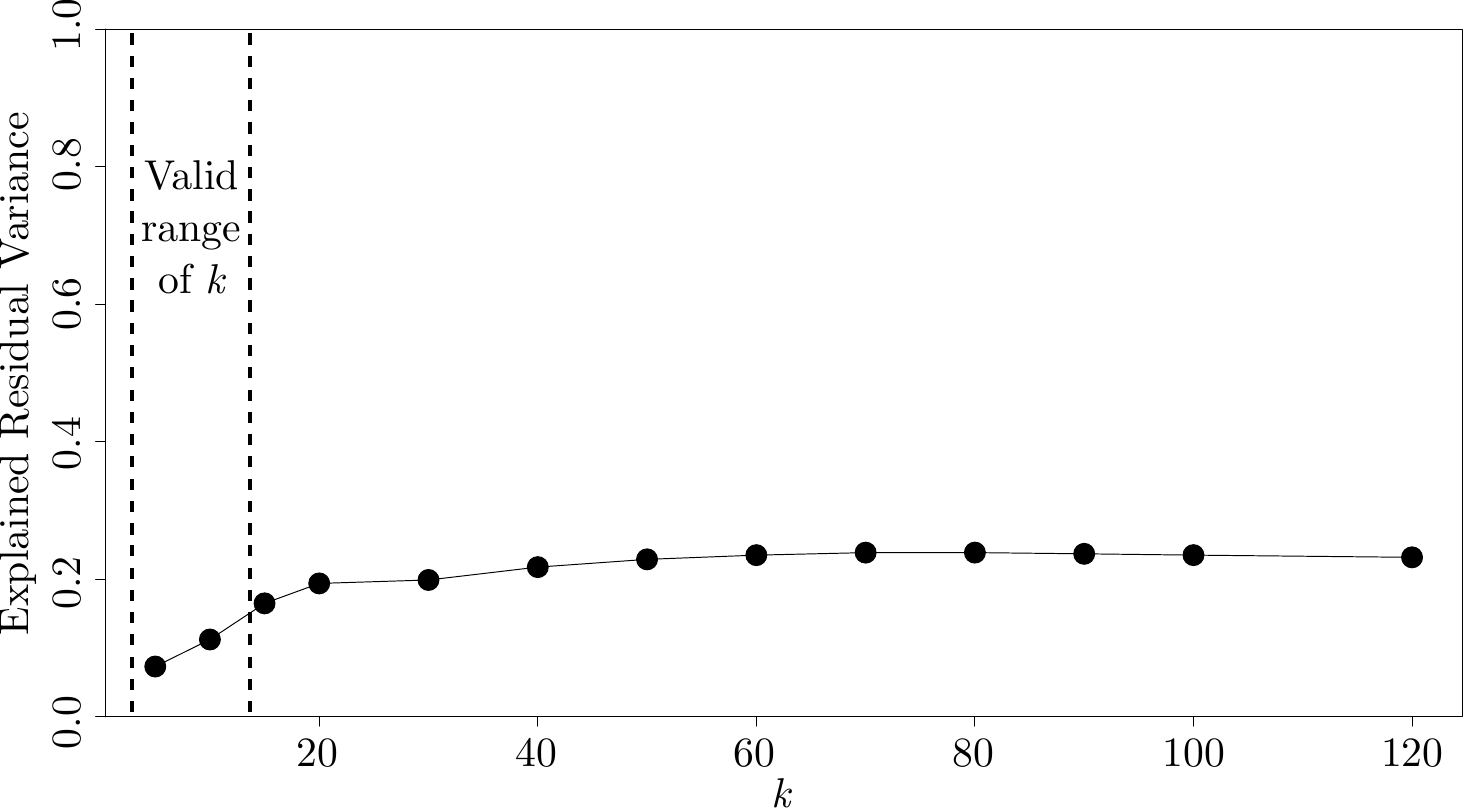}}
  \caption{Residual variance of Isomap embedding to 3 dimensional space vs different Isomap $k$, for pinball dataset $Re = 130$ with the valid range of Isomap's $k$ according to \cite{valid_range_k}}
\label{fig:res_var_comparison}
\end{figure}

 Different approaches have been proposed in the literature to cope with the selection of $k$ in Isomap. In this work, we follow the methodology presented by \citet{valid_range_k} to determine a valid range of values $[k_{\min},k_{\max}]$ to perform the search. The lower bound of the interval, $k_{\min},$ is selected as the smallest $k$ so that the neighboring graph $G$ is connected. Regarding the upper bound, $k_{\max},$ pick the largest value of $k$ so that the following equation holds:
\begin{equation} \label{eq:kmax}
    \displaystyle\frac{2E}{N} \leq k + 2,
\end{equation}
where $E$ is the number of edges  and $N$ is the number of nodes in $G$. Once the valid range of $k$ is found, \citet{valid_range_k} propose to pick $k\in[k_{\min},k_{\max}]$ so that the residual variance is minimum.
In all our studies in section \ref{sec:results}  using any value in the  valid range selected this way  results in a low residual variance. We refer the reader to Figure \ref{fig:res_var_comparison} for an illustration of the impact of the choice of $k$ on the residual variance for the pinball dataset for $Re = 130.$

\section{Different approaches to define the residual variance.}\label{app:residualvariance}

There exist numerous approaches to assess the fits provided by different dimensionality reduction techniques. Regarding Isomap, the so-called residual variance as stated in \eqref{eq:residualvarianceT}  is the common choice. 

Regarding POD, the classical definition of residual to assess its performance is given by

\begin{equation}\label{eq:residualvariancePODclassical}
   1 - \displaystyle\frac{\displaystyle\sum_{j = 1}^p \text{var} (\mathbf{z}_j)}{\displaystyle\sum_{j = 1}^P \text{var} (\mathbf{z}_j)},
\end{equation}
where var states for the variance and $\mathbf{z}_j$ are the Principal Components, $j=1,\ldots,P.$

The ways of measuring the fits in Isomap and POD given by \eqref{eq:residualvarianceT} and \eqref{eq:residualvariancePODclassical} are not comparable. In order to evaluate Isomap and POD up to the same standard, \cite{Tenenbaum2319} proposed to replace the geodesic distances approximated by $\mathbf{D}_{G}$ in \eqref{eq:residualvarianceT} by the pairwise Euclidean distances in the input high-dimensional space, $\mathbf{D}_{\mathbf{X}}$, where the element in row $i$ and column $j$ in $\mathbf{D}_{\mathbf{X}}$ is $\Vert\mathbf{x}_i - \mathbf{x}_j \Vert_2,$ $i,j=1,\ldots,n.$ Then a related definition of residual variance for POD is,

\begin{equation}\label{eq:residualvariancePODT}
    1 - R^2(\text{vec}(\mathbf{D}_{\mathbf{X}}), \text{vec}(\mathbf{D}_{\mathbf{Z}})),
\end{equation}
where $\mathbf{D}_{\mathbf{Z}}$ is the matrix of Euclidean distances between the retained Principal Components $\mathbf{z}_1,\ldots, \mathbf{z}_d$.  

Nevertheless, the definition of residual variance in \eqref{eq:residualvariancePODT} does not capture the ability of POD of reproducing the geodesic distances in the high-dimensional space. To do so, a different but related proposal to measure the residual variance for POD combines \eqref{eq:residualvarianceT} and \eqref{eq:residualvariancePODT} as
\begin{equation}\label{eq:residualvariancePODIaniro}
    1 - R^2(\text{vec}(\mathbf{D}_{G}), \text{vec}(\mathbf{D}_{\mathbf{Z}})).
\end{equation}

Finally, we point out that the residual variance as defined in \eqref{eq:residualvarianceT}, \eqref{eq:residualvariancePODT} and  \eqref{eq:residualvariancePODIaniro} may not decrease strictly when the number of dimensions increases. In other words, the correlation between a distance matrix obtained from a set of points in $\mathbb{R}^d$, $\mathbf{D_{d}}$, and another distance matrix, $\mathbf{D}$, does not necessarily decrease if the distance between the embedded points in  $\mathbb{R}^{d-1}$ by dropping one of the dimensions, $\mathbf{D}_{d-1}$, is considered instead.

\begin{figure}
     \centering
     \begin{subfigure}[b]{0.49\textwidth}
         \centering
         \includegraphics[width=\textwidth]{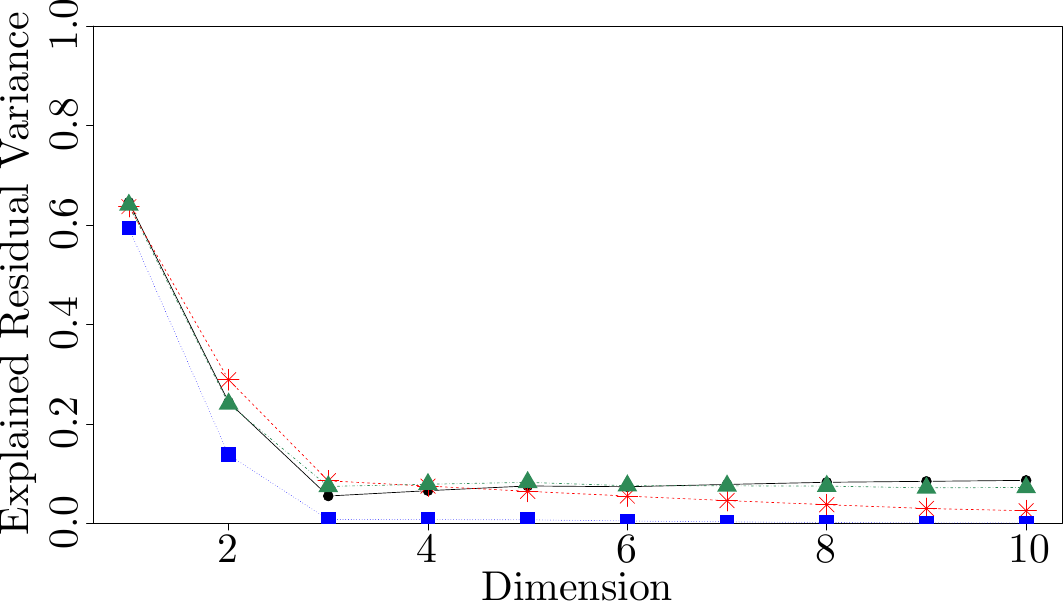}
         \caption{}
         \label{fig:all_rev_var_Re80}
     \end{subfigure}
     \hfill
     \begin{subfigure}[b]{0.49\textwidth}
         \centering
         \includegraphics[width=\textwidth]{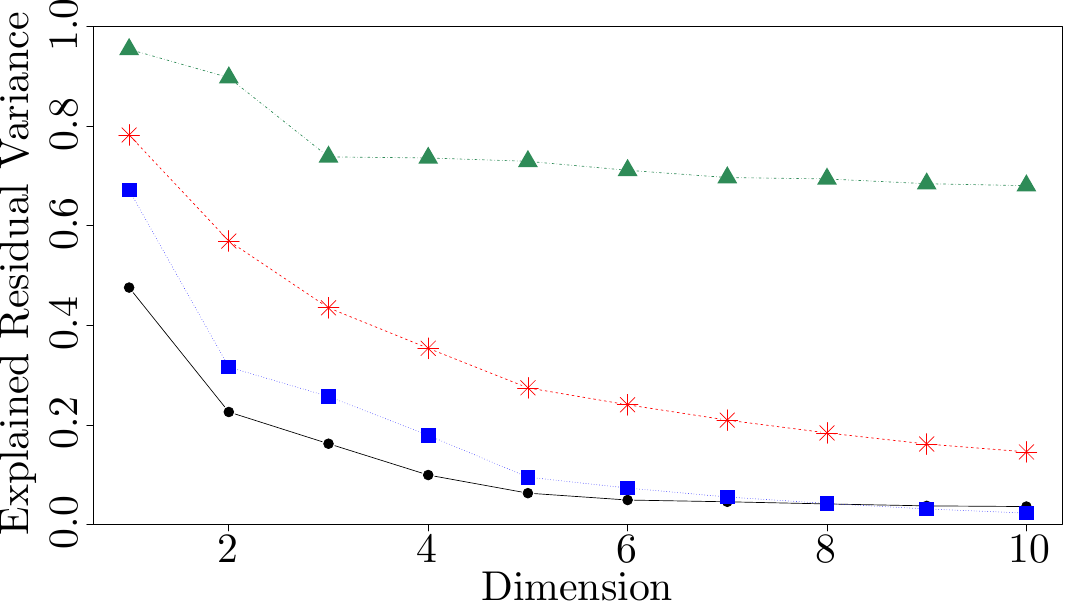}
         \caption{}
         \label{fig:all_rev_var_Re130}
     \end{subfigure}
     \hfill
     \begin{subfigure}[b]{0.49\textwidth}
         \centering
         \includegraphics[width=\textwidth]{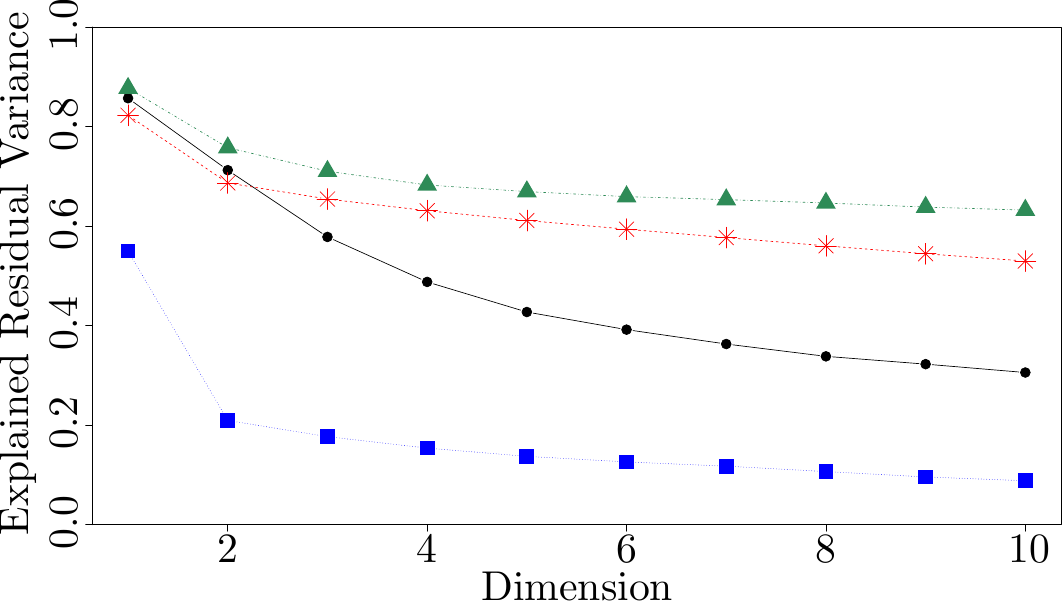}
         \caption{}
         \label{fig:all_rev_var_SwJ}
     \end{subfigure}
     \hfill
     \begin{subfigure}[b]{0.49\textwidth}
         \centering
         \includegraphics[width=\textwidth]{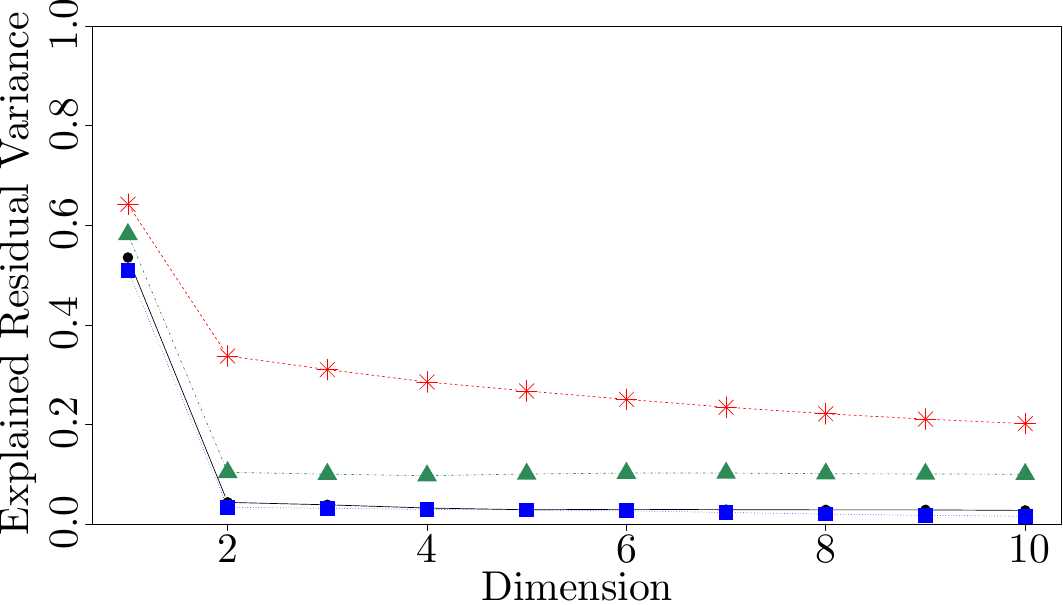}
         \caption{}
         \label{fig:all_rev_var_tandem}
     \end{subfigure}
        \caption{Different methods for defining residual variance. a) Pinball- Re80- b) Pinball- Re130- c) Swirling Jet- d) Tandem Cylinders; Black: \eqref{eq:residualvarianceT}- Red: POD \eqref{eq:residualvariancePODclassical}- Blue: POD \eqref{eq:residualvariancePODT}- Green: POD \eqref{eq:residualvariancePODIaniro}}
        \label{fig:all_rev_var}
\end{figure}

Figure \ref{fig:all_rev_var} shows the values for the different definitions of residual variances presented in this work in each  of the case studies in section \ref{sec:results} for different numbers of dimensions.  The black line corresponds to \eqref{eq:residualvarianceT}, the red one is \eqref{eq:residualvariancePODclassical}, blue \eqref{eq:residualvariancePODT} and green \eqref{eq:residualvariancePODIaniro}.
For all the dimensions considered, the residual variance of Isomap surpasses POD using definitions \eqref{eq:residualvarianceT} and \eqref{eq:residualvariancePODT}.

\bibliographystyle{plainnat}
\bibliography{jfm}

\end{document}